\documentclass[twocolumn]{aastex62}

\usepackage{bm}

\newcommand{\FuncCall}[2]{\texttt{\bfseries #1(#2)}}
\newcommand{\assign}{\leftarrow}

\usepackage{hyperref}
\usepackage[linesnumbered,ruled]{algorithm2e}

\graphicspath{{./}{figures/}}

\accepted{July 20, 2021}

\submitjournal{AJ}

\shortauthors{Yip et al.}

\begin{document}

\title{Peeking inside the Black Box:\\Interpreting Deep Learning Models for Exoplanet Atmospheric Retrievals}

\correspondingauthor{K.H. Yip}
\email{kai.yip.13@ucl.ac.uk}

\author[0000-0002-9616-1524]{Kai Hou Yip}
\affil{Department of Physics and Astronomy \\
		University College London \\
		Gower Street,WC1E 6BT London, United Kingdom}

\author[0000-0001-6516-4493]{Quentin Changeat}
\affiliation{Department of Physics and Astronomy \\
		University College London \\
		Gower Street,WC1E 6BT London, United Kingdom}

\author[0000-0001-8453-7574]{Nikolaos Nikolaou}
\affiliation{Department of Physics and Astronomy \\
		University College London \\
		Gower Street,WC1E 6BT London, United Kingdom}

\author[0000-0001-8587-2112]{Mario Morvan}
\affiliation{Department of Physics and Astronomy \\
		University College London \\
		Gower Street,WC1E 6BT London, United Kingdom}
		
\author[0000-0002-5494-3237]{Billy Edwards}
\affiliation{Department of Physics and Astronomy \\
		University College London \\
		Gower Street,WC1E 6BT London, United Kingdom}
		
\author[0000-0002-4205-5267]{Ingo P. Waldmann}
\affiliation{Department of Physics and Astronomy \\
		University College London \\
		Gower Street,WC1E 6BT London, United Kingdom}

\author[0000-0001-6058-6654]{Giovanna Tinetti}
\affiliation{Department of Physics and Astronomy \\
		University College London \\
		Gower Street,WC1E 6BT London, United Kingdom}

\begin{abstract}
Deep learning algorithms are growing in popularity in the field of exoplanetary science due to their ability to model highly non-linear relations and solve interesting problems in a data-driven manner. Several works have attempted to perform fast retrievals of atmospheric parameters with the use of machine learning algorithms like deep neural networks (DNNs).  Yet, despite their high predictive power,  DNNs are also infamous for being `black boxes’. It is their apparent lack of explainability that makes the astrophysics community reluctant to adopt them. What are their predictions based on? How confident should we be in them? When are they wrong and how wrong can they be? In this work, we present a number of general evaluation methodologies that can be applied to any trained model and answer questions like these.  In particular, we train three different popular DNN architectures to retrieve atmospheric parameters from exoplanet spectra and show that all three achieve good predictive performance. We then present an extensive analysis of the predictions of DNNs, which can inform us —among other things — of the credibility limits for atmospheric parameters for a given instrument and model. Finally, we perform a perturbation-based sensitivity analysis to identify to which features of the spectrum the outcome of the retrieval is most sensitive. We conclude that for different molecules, the wavelength ranges to which the DNN’s predictions are most sensitive, indeed coincide with their characteristic absorption regions. The methodologies presented in this work help to improve the evaluation of DNNs and to grant interpretability to their predictions.
\end{abstract}


\section{Introduction} \label{sec:intro}

Exoplanetary science is one of the fastest expanding fields in astronomy. The increasing number of discovered exoplanets provided necessary motivation for subsidiary disciplines to grow. 

Exoplanet atmosphere characterisation, in particular, is one of the frontiers of the field. Transit spectroscopy, which consists in observing transits at different wavelengths has allowed astronomers to robustly detect chemical species such as water vapour, carbon-bearing molecules, oxides and alkali species in the atmosphere \citep{Linsky2010, Fossati2010,berta2012, mandell2013,dekok2013, Ehrenreich2014,Barman2015,Macintosh2015,Macdonald2017,Jacob2018,edwards_w76}. These successes are built upon the foundation of generations of ground based and space based instruments, such as the Very Large Telescope (VLT), the Spitzer Space Telescope and the Hubble Space Telescope. The accumulation of such observations over the years has enabled large scale statistical studies of subpopulations of exoplanets, for example on a wide range of Hot Jupiter atmospheres \citep{sing2016,iyer_pop,population,fisher_18}.

Looking forward, the next generation of space missions, dedicated to exoplanet characterisation, such as Ariel \citep{ARIEL}, Twinkle \citep{edwards_exo} and JWST \citep{JWST_mission}, will be launching within the next decade, delivering spectra with broader wavelength coverage and higher spectral resolution. The prospect of better data quality has encouraged further development in forward modelling of exoplanet spectra and atmospheric retrieval techniques \citep[e.g.][]{Irwin_nemesis,Madhusudhan_retrieval,Line_chimera,Ormel_arcis,Zhang_platon,Taurex3}. 

Artificial intelligence and in particular deep learning, has risen in popularity in recent years. Deep learning algorithms have proven successful in efficiently deriving useful models from large amounts of high-dimensional data. Such models, capable of capturing highly non-linear relationships have been used to solve hard problems in a wide range of application domains, such as image classification \citep{ImageReview}, natural language processing \citep{NLPreview} and time series analysis \citep{TSCreview}. Typically, training such complex models without overfitting requires a large amount of data.

The use of deep learning (DL) and general machine learning (ML) algorithms has become widespread in the field of exoplanet research. \cite{McCauliff2015} applied Random Forests (RFs) to identify candidate transit signals from Kepler light-curves. Subsequently \cite{Shallue_2018} and \cite{Pearson2018} demonstrated the potential of DL in transit candidate vetting and inspired followup applications on other instruments \citep[e.g.][]{Alexander_2019,Dattilo2019,Osborn2020,Yu2019}. \cite{Schanche2019} developed a combination of shallow ML and DL models to improve vetting accuracy on WASP data. Additionally, Long Short-Term Memory Network \citep{Mario2020} and Random Forest \citep{Krick2020} were implemented to model and correct the systematics of Spitzer IRAC exoplanet transit modelling and detection.

On the planetary characterisation front, despite the fact that retrieval has always been the standard, universal approach when it comes to inferring atmospheric properties, retrieval frameworks are not without weaknesses. It is essentially a fitting algorithm that attempts to estimate a best-fit solution given a forward model and a list of parameters with their bounds. The limitation imposed by the observational data means that multiple solutions, regardless of their feasibility, could exist, and it is left for the user to judge the feasibility of the outcome. A neural network, on the other hand, is able to learn the intricate, non-linear relationships between parameters and the observational data. The development of a neural network driven retrieval is still at its early stages, but there have been attempts to infer atmospheric properties from a network. \cite{robert_ingo} pioneered the application of deep learning models to identify the existence of molecular species in a transmission spectrum. \cite{Marquez2018} used RFs to infer atmospheric properties, such as temperature and water abundance, from exoplanet spectra. The success of RFs inspired further applications. \cite{fisher2020} applied the same algorithm on high resolution ground-based observations. \cite{Nixon2020} built upon \cite{Marquez2018}'s work and produced an RF generated posterior distribution with excellent agreement to one from a fully Bayesian retrieval. On the neural network front, \cite{exogan} utilised a Generative Adversarial Network (GAN), a deep learning network architecture that can generate the closest synthetic spectrum and its associated atmospheric properties to a given observed spectrum. \cite{Cobb2019} developed a Bayesian Neural Network to model the posterior distribution between atmospheric parameters. To speed up the computationally expensive radiative transfer simulation process, \cite{Himes2020} trained a ML surrogate forward model and demonstrated its potential to significantly reduce retrieval time.

However, despite their predictive power, models generated using some of the most powerful machine learning algorithms --deep learning being the most prominent example-- are often regarded as `black boxes'. Deep neural networks trained on large, high-dimensional datasets are models that typically contain thousands or even millions of parameters learnt from data. This is what allows them to model complex non-linearities within the data and ultimately leads to their accurate predictions. But the same complexity also makes it difficult to understand what factors contribute the most to DNN predictions. Developing methodologies to make such models more interpretable is a growing area in the field of machine learning \citep{molnar2019}. Better interpretability and a more robust understanding of the DNN's uncertainties may lead to a broader adoption of these methods in the physical sciences. It allows us to understand if our models make correct predictions for the right reasons, why our models are wrong --when they are-- and how to correct for their biases. Beyond this, interpretability methods allow for identifying the ever-present biases in the dataset -- especially relevant in the case of astrophysics where a lot of effort is dedicated to analysing simulated data in preparation for deploying a new instrument. Finally, understanding machine learning models in terms we can relate to the underlying theory of the application domain (e.g. astrophysics), can provide us with new theoretical insights. 

In this paper, we will investigate the use of several DNN architectures (MLPs\footnote{Multilayer Perceptrons (MLPs) are fully (i.e. `densely') connected feed-forward neural networks. They are the oldest type of DNN developed and the most commonly used in structured data.}, CNNs, LSTMs) in the problem of exoplanet atmospheric retrievals. We emphasise that our goal is not to train a neural network to perform retrieval, the two methodologies might have similar outcomes but they are different in their approach. Our goal is to investigate how to probe into the inner workings of DNN models trained to perform this prediction task. We will demonstrate how to analyse the performance of a trained model and answer questions like: `What is the true abundance range if the model predicts an abundance of H$_2$O of $10^{-5}$?' We will use this analysis to explore the performance of an instrument/observational strategy (in our case the Deep survey by the ESA Ariel space telescope \citep{ARIEL}). Moreover, we will present a general methodology that can be applied to any trained neural network (or rather any statistical predictive model) to understand how its input affects its predictions. The proposed method is a quantification of the sensitivity of the trained model to the various features of the input. In the context of atmospheric retrievals, we will visualise how different features of the spectrum affect the quality of the retrieval. In other words we ask: `Where does a neural network look in the spectrum to determine the value of each of the retrieved parameters?' As we will see, the answer mostly agrees with our physical intuition, yet it occasionally brings to light interesting new insights about the model, the data, or the underlying physics. Our implementation is available on Github\footnote{\url{https://github.com/ucl-exoplanets/Spectra_Sensitivity_analysis}} and Zenodo \citep{khy_sensitivity}.



\section{Problem Statement, Data \& Models}
\subsection{Objectives}
There are three main objectives in this investigation:
\begin{enumerate}
  \item To train DNNs to infer different atmospheric parameters from a transmission spectrum. We demonstrate that several DNN architectures (MLPs, CNNs, LSTMs) are capable of producing models that achieve good predictive performance in this task. We use the best model obtained at this stage as an example model for the next two stages (which, we should note, are not tied to any specific model or learning algorithm).

\item To present a detailed evaluation methodology to investigate the quality of the predictions of any given trained model. We move beyond the naive regression visualisations and demonstrate how to decompose the error of the model into its bias and variance components, how to check for interactions among variables and how to assess the credibility of its predictions. In doing so, we also infer the limits of credibility on each target on the given dataset under our model.

\item To introduce a perturbation-based sensitivity analysis approach for visualising regions of the input that are most relevant for the predictions of any given trained model. Doing so, allows us to understand whether the regions of the input the model is most sensitive to, align with our physical intuition. 
\end{enumerate}

\subsection{Data Generation}
\label{sec:gen_data}
\begin{table}[]
\resizebox{\columnwidth}{!}{%
\begin{tabular}{c|ccc}
AMPs & Range & Scale & Sampling \\ \hline
H$_2$O & -9 to -3 & log & Uniform \\
CH$_4$ & -9 to -3 & log & Uniform \\
CO & -9 to -3 & log & Uniform \\
CO$_2$ & -9 to -3 & log & Uniform \\
NH$_3$ & -9 to -3 & log & Uniform \\
M$_p$ {[}log(M$_J$){]} & -3.00 to 1.43 & log & \cite{target_list} \\
R$_p$ {[}R$_J${]} & 0.07 to 2.39 & linear & \cite{target_list}  \\
T$_p$ {[}K{]} & 1393 to 3999 & linear & \cite{target_list} \\
Cloud {[}log(Pa){]} & 2.7 to 6 & log & Uniform
\end{tabular}
}
\caption{Sampling range, scale and sampling method used for different AMPs in the synthetic dataset.}
\label{tab:prior}

\end{table}
For the purposes of this study, we generated synthetic\footnote{Although the specific models trained on this dataset, their predictive performance and sensitivity analysis results are problem-specific, it is important to clarify that all methodologies presented in this work are applicable to any machine learning model trained on a given dataset.} planetary atmospheres from planets contained in the Ariel Target list \citep{target_list}.  A total of 11940 transmission spectra were produced. A transmission spectrum records the $\lambda$ dependency change in transit depth ($\Delta$t$_\lambda$). This large scale spectrum generation is made possible through the function \textit{Alfnoor-forward} in Alfnoor \citep{alfnoor}, a pipeline consisting of TauREx3 and ArielRad, the Ariel Radiometric Model \citep{mugnai_Arielrad}. Each generated spectrum is binned to Ariel Tier-2 resolution with error-bars calculated based on Deep survey requirements and realistic estimates of the instrument, observations taken and planet observed. The Ariel Tier 2 resolution is kept the same throughout the investigation, any binning process is not performed in wavelength space. This setup was used by \cite{alfnoor} in their investigation, which provided a benchmark for us to compare in Section \ref{sseq:credibility}. Here we denote the binned spectrum as the mean, ground truth spectrum $\bar{\mathbf{X}}$ = [$\bar{x}_1$, $\bar{x}_2$, \dots, $\bar{x}_{52}$], where $x_i$ represents the transit depth at the $i$-th wavelength bin in ascending order. The associated uncertainty for each $x_i$ will be denoted as $\sigma_i$. 

All generated spectra are subject to the same assumptions: the atmosphere for each spectrum is assumed to have constant He/H$_2$ ratio of 0.17, a hypothesis corresponding to a primary atmosphere with solar composition. Rayleigh scattering and Collision Induced Absorption for H$_2$-H$_2$ and H$_2$-He are included. The T-P profile is assumed to be isothermal and trace gases are introduced to the atmosphere with iso-abundance profiles (profiles constant with altitude). Other planetary parameters necessary in producing a transmission spectrum and estimating the observational uncertainties (spectrum and error bars), such as stellar radius (R$_s$), planet radius (R$_p$), planet mass (M$_p$), planet temperature (T$_p$) and other orbital parameters (semi-major axis, distance to the star, eccentricity) are taken from the predictions in \cite{target_list}.

To generate an unbiased sample of spectra, we added a number of trace gases. For each constituent trace gases (H$_2$O, CH$_4$, CO, CO$_2$ and NH$_3$), we uniformly sampled their log abundance from -9 to -3\footnote{The range was chosen to explore Ariel Deep survey's ability at capturing low molecular abundances. Log abundance values $> -3$ are omitted as they can easily be detected via current retrieval methods. }. The line lists of different molecules are taken from ExoMol \citep{Tennyson_exomol}, HITRAN \citep{gordon} and HITEMP \citep{rothman}. Additionally, we have also added grey clouds with its cloud deck pressure log(P$_{cloud}$) uniformly sampled from 2.7 to 6. Table \ref{tab:prior} summarises the sampling range, sampling method and their respective scales.  For a detailed discussion of the data generation process, we  refer the interested readers to Section 2.2 of \cite{alfnoor}.

\subsection{Data Preprocessing}
\label{sec:preprocessing}
The term `parameter' is defined differently under the context of machine learning and exoplanet atmospheric retrievals. To explicitly distinguish the different contexts, we will refer to Atmospheric Model Parameters as `AMPs' and Deep Neural Network parameters (synaptic weights) as `weights' hereafter. 

\begin{figure}
    \centering
    \includegraphics[width=\columnwidth]{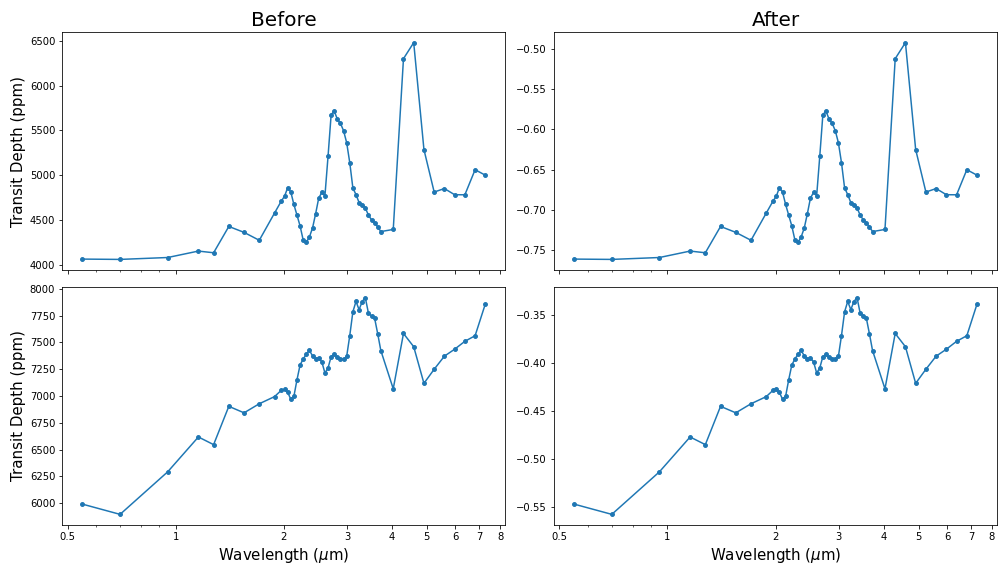}
    \caption{Empirical comparison of a sample of synthetic spectra before and after standardisation. The comparison shows that our transformation only scales the spectral features, without distorting their relative shape.}
    \label{fig:spectra_compare}
\end{figure}

The synthetic spectra and their corresponding AMPs are standardised (normalised so that each feature, i.e. wavelength bin and each AMP has zero mean and unit variance) to facilitate the training of the DNN models (see Figure \ref{fig:spectra_compare} for empirical comparison of sample spectra before and after standardisation). The standardised dataset is then split uniformly at random into three subsets, the original training set (70\%), the validation set (10\%) and the test set (20\%).
The original training set is not directly used in training. Rather, it is used to generate an augmented training set. For each datapoint (spectrum) $\bar{\mathbf{X}}$ in the original training set, we generate 50 datapoints $\tilde{\mathcal{X}}$ for the augmented training set. Each $\tilde{\mathcal{X}}$ is produced by sampling a new $\tilde{x_i}$ from a Gaussian distribution centered at $\bar{x_i}$ and having a standard deviation defined by the corresponding $\sigma_i$. The original ground-truth (i.e. noise-free) spectra are thus discarded and the models are trained only on these noisy, more realistic samples. The same applies for the validation set, whereas the test set is kept noise-free.

\subsection{Model Training}
\label{sec:training}
 We trained a DNN to perform a multi-output regression task. The task is to predict nine targets, log(X$_{H_2O}$), log(X$_{CH_4}$), log(X$_{CO}$), log(X$_{CO_2}$), log(X$_{NH_3}$), R$_p$, log(M$_p$), T$_p$ and log(P$_{cloud}$) from a given spectrum. For ease of referencing we denote the AMPs as $\mathbf{y} = [y_1,y_2, ... y_9]$, where $y_j$ represents the $j$-th AMP (as ordered above). The model is trained in a supervised manner by minimising the Mean Squared Error (MSE) between the predicted values $\mathbf{\hat{y}}$ and the ground truth $\mathbf{y}$, averaged across all targets. Details of how the neural networks were trained can be found in Appendix \ref{app:imple}. The results and figures shown in this paper are selected from our best performing model, an 1-dimensional Convolutional Neural Network (1D-CNN). 
\section{Evaluation of Predictive Models} \label{sec:method}

\subsection{Prediction versus Truth Plot}
\begin{figure}
    \centering
    \includegraphics[width=\columnwidth]{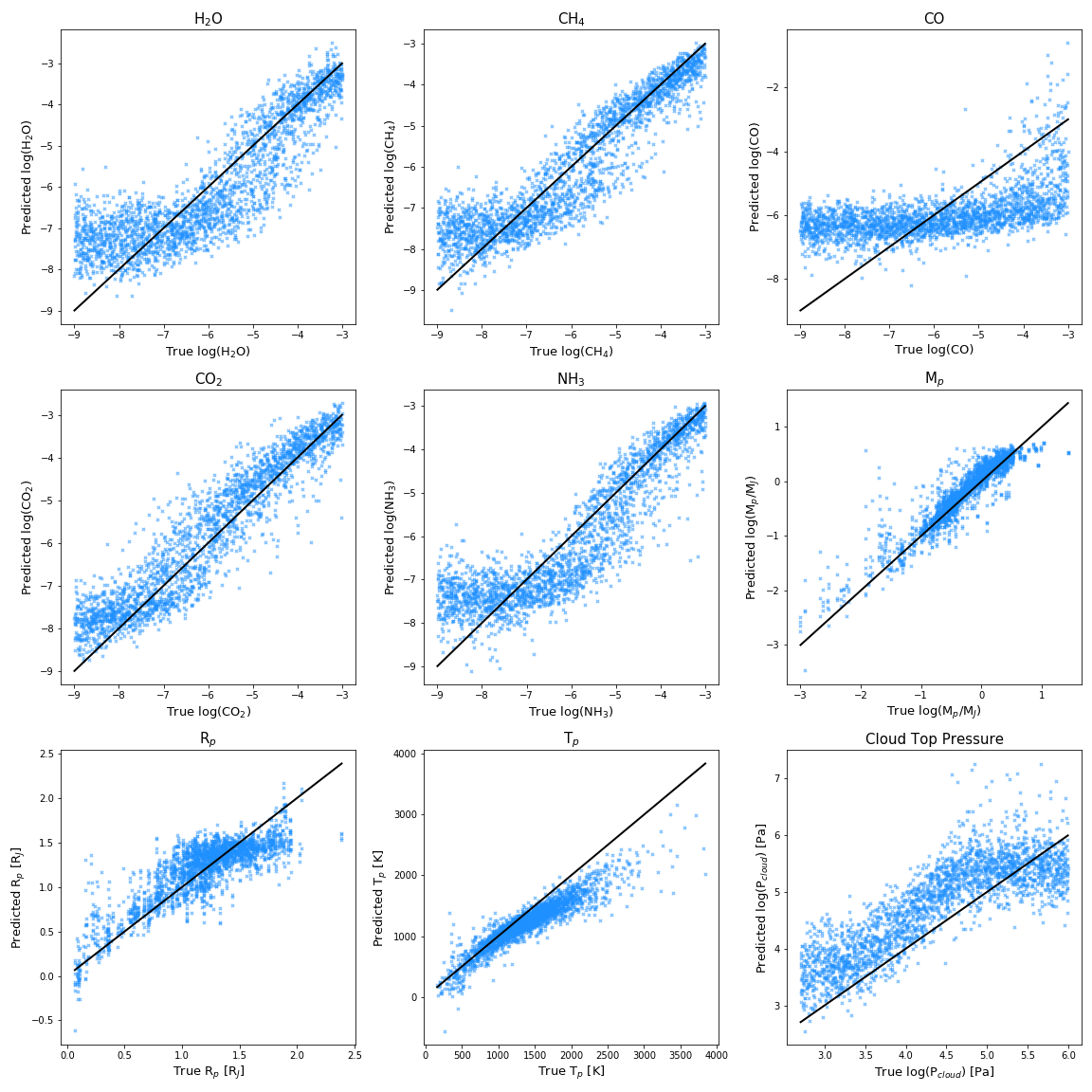}
    \caption{Prediction versus Truth Plot for each AMP in their respective units. Each blue point represents a single spectrum from the test set. Prediction from the model (y-axis) is plotted against the corresponding ground truth (x-axis). The black diagonal line represents the distribution of perfect predictions. }
    \label{fig:all_distributions}
\end{figure}
In Figure \ref{fig:all_distributions} we compare for each of the individual AMPs to be retrieved, the value predicted by the model (y-axis) against the true value (x-axis). These predictions were generated for noise-free spectra from the test set. Each blue point in every subplot represents prediction from a single (test) spectrum. The diagonal line represents the predictions of a perfect model (one that always predicts the ground truth). This is a classic visualisation of regression results. It is useful for obtaining an overall sense of the model's performance: ideally `points should not deviate much from the diagonal'. But significant information is obscured by the fact that (i) the density of points is not uniform and that (ii) `deviation' can assume different mathematical meanings (e.g. `average deviation', `standard deviation around the mean').

\subsection{Bias \& Variance Visualization}
\label{sec:BVPlot}

\begin{figure*}
    \centering
    \includegraphics[width=\textwidth]{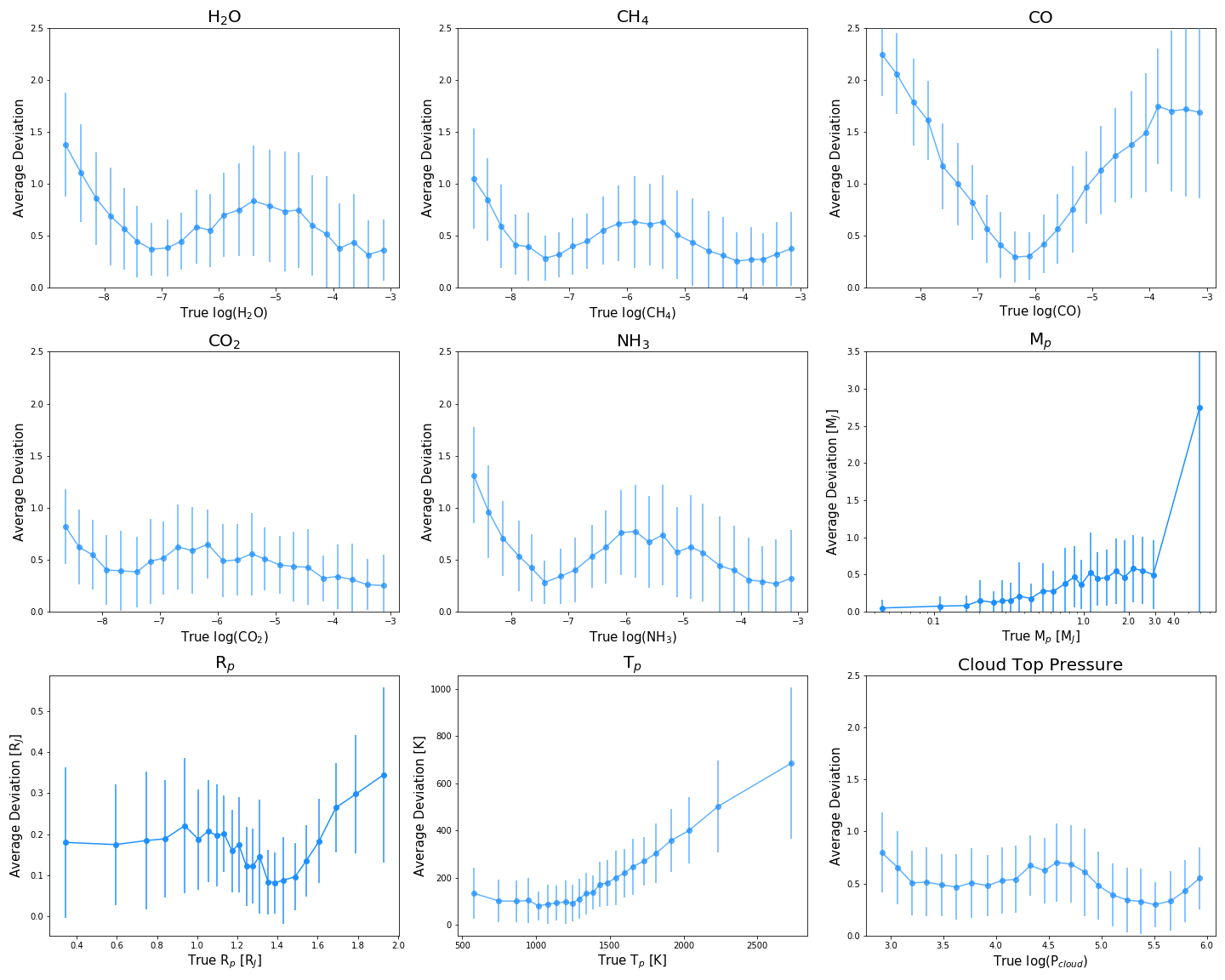}
    \caption{Visualization of bias \& variance for different AMPs. Each point represents the average absolute deviation of the model's prediction from the ground truth (a measure of the model's bias) and the associated error-bar represents the 1-$\sigma$ spread of the predictions (a measure of the model's variance). Note that the model's variance also includes contributions from the irreducible noise. }
    \label{fig:BVPlot}
\end{figure*}

To get a deeper understanding of the model's performance, we need to go beyond Figure \ref{fig:all_distributions}. The error (in our case, the MSE) of a regression model can be decomposed into 3 terms: bias, variance and irreducible noise \citep[][]{Murphy_2012}. Here we state explicitly our definition of bias and variance to avoid any confusion with the terminologies. Bias refers to the mean absolute deviation of the model's predictions from the true value. Variance refers to the spread/width of the model’s prediction. Irreducible Noise (third term) refers to variance inherent to the data\footnote{The noise component is due to the inherent uncertainty in predicting the targets (AMPs) from the data (spectra), even with a `perfect' model. As such it is irreducible.}. We note here that the variance we computed for each wavelength bin in the following figures will inherently contain variance from both model’s prediction and the irreducible noise.

An approximate visualisation of the bias and variance components of the error for each AMP as shown in Figure \ref{fig:BVPlot} \footnote{The interested reader can find the same type of plots generated by the other DNN models examined in Figures \ref{fig:mlp_bvp} \& \ref{fig:lstm_bvp}.} can be more illuminating. Each subplot is generated by equal frequency binning of the true values (each bin contains 100 data points). For each bin we calculate the mean absolute difference between pairs of predicted and true values $|\hat{y}_j - y_j|$ (a proxy of the model's bias) and the standard deviation of these differences (a proxy of its variance and variance from irreducible noise ). Ideally, one would like to keep both components of the error low, i.e. to consistently predict values close to the truth. Using Figure \ref{fig:BVPlot} we can inspect regions where the model's predictions suffer from high bias (deviation), high variance (spread) or both. 

In our application on the simulated Ariel-like dataset, the model's predictions on the gases exhibit a similar trend: the prediction starts off with small bias and variance at high abundances, both the bias and the variance gradually become higher as the abundance drops, reaching a peak at certain abundance. At lower abundances, below the credibility limit (see Section~\ref{sseq:credibility}) of the corresponding gas, the network resorts to --on average-- outputing an  `average low' value. This results in a characteristic trough (its minimum indicating the `average low' value for each gas). This behaviour is expected. A molecule's absorption feature is most prominent at high abundances and this helps to tightly constrain the model's predictions. However, as gas abundance decreases, so does the magnitude of the corresponding feature, making it easier for other absorption features to partially, or --in some cases-- even completely mask it. The task therefore becomes progressively harder, which contributes to a higher variance and a significantly biased mean deviation. If the level of abundance becomes low enough, the model can no longer constrain the prediction due to presence of features from other molecules. At this point, the best strategy for a loss minimising model is to restrict its output and output a limited range of values centred at an average value in the low abundance region (see discussion in Section \ref{sseq:credibility}).

The above trend is generally followed by most gases except CO, which has most of its predictions clustered around log(X) = -6 exhibiting large bias. The poor quality of the predictions for CO is expected, given the spectral coverage of the instrument and lack of broad-band features from the molecule itself. The lack of information on CO causes the model  to minimise the loss by always predicting a restricted range of values with an average volume mixing ratio of log(X) = -6  (see Figure \ref{fig:deivation}). 

On the other hand, for planetary parameters such as M$_p$, R$_p$ and T$_p$, the performance of the model varies. The model's M$_p$ predictions are generally characterised by low bias. The ability of the model to predict M$_p$ accurately, suggests that the Ariel Tier 2 spectra alone contain sufficient information to constrain M$_p$, confirming the findings in \cite{quentin_mass}. The model's performance on R$_p$ and T$_p$, however, is not as satisfactory. While the model is able to accurately predict smaller planetary radii, its predictions on the very largest radii in the sample are characterised by high bias. In particular, the radius is consistently underestimated. For T$_p$, the model becomes progressively more biased in hotter temperatures starting from T$_p$ = 1500K, again consistently underestimating them (see Figure \ref{fig:all_distributions}). 

Regarding the model's prediction of log(P$_{cloud}$), it appears that most of the predictions have been stratified into two levels, a sign that the model is only able tell qualitatively whether the atmosphere is cloudy or not. 

The poor performance in R$_p$, T$_p$ and log(P$_{cloud}$) can be explained by several factors. Most notably:
\begin{itemize}
    \item [1.] The degeneracy involving these quantities. It is well known that the interaction between these quantities could produce very similar spectral features \citep[e.g.][]{Brown_2001,deWit_mass, Griffith_degen_2014,fortney_clouds,fisher_18,Rocchetto_biais_JWST,Lecavelier_des_Etangs_2008,Tinetti_ariel,quentin_mass}. For example, the model tends to underestimate T$_p$ and R$_p$ and overestimate log(P$_{cloud}$) (i.e. predict a less cloudy atmosphere). All of these AMPs can compensate each other producing similar spectra. As there are more than one possible solutions, our  model (being a deterministic function outputting a single prediction for each AMP) fails to always identify the `true' (i.e. in the data generation sense) solution given the limited information from the data. Ideally, in an atmospheric retrieval setting, rather than predicting the most probable values of the AMPs, we would rather predict their posterior distribution or at least capture their covariance. In Section \ref{sec:interaction} we perform an initial investigation of interactions among AMPs.

    \item [2.] The standardising step in Section \ref{sec:preprocessing} was performed by extracting the overall mean and standard deviation of the training set, the order of magnitude differences between spectra may have significantly reduced the dynamic range within a spectrum, dwarfing any molecular signatures.   
    \item [3.] The non-uniform distribution of R$_p$, T$_p$ and M$_p$ in the generated data. The non-uniformity means that the model will focus on more accurately predicting values in the densely-populated areas of the target space to the potential detriment of the quality of its predictions elsewhere, if that means achieving a lower MSE \footnote{The non-uniform distribution could be alleviated with better knowledge on permitted combinations of R$_p$, T$_p$ and M$_p$, an alternative way is to adjust the weight of each sample based on its rarity (heavier loss on uncommon examples). }. 
\end{itemize}

\begin{figure*}
    \centering
    \includegraphics[width=\textwidth]{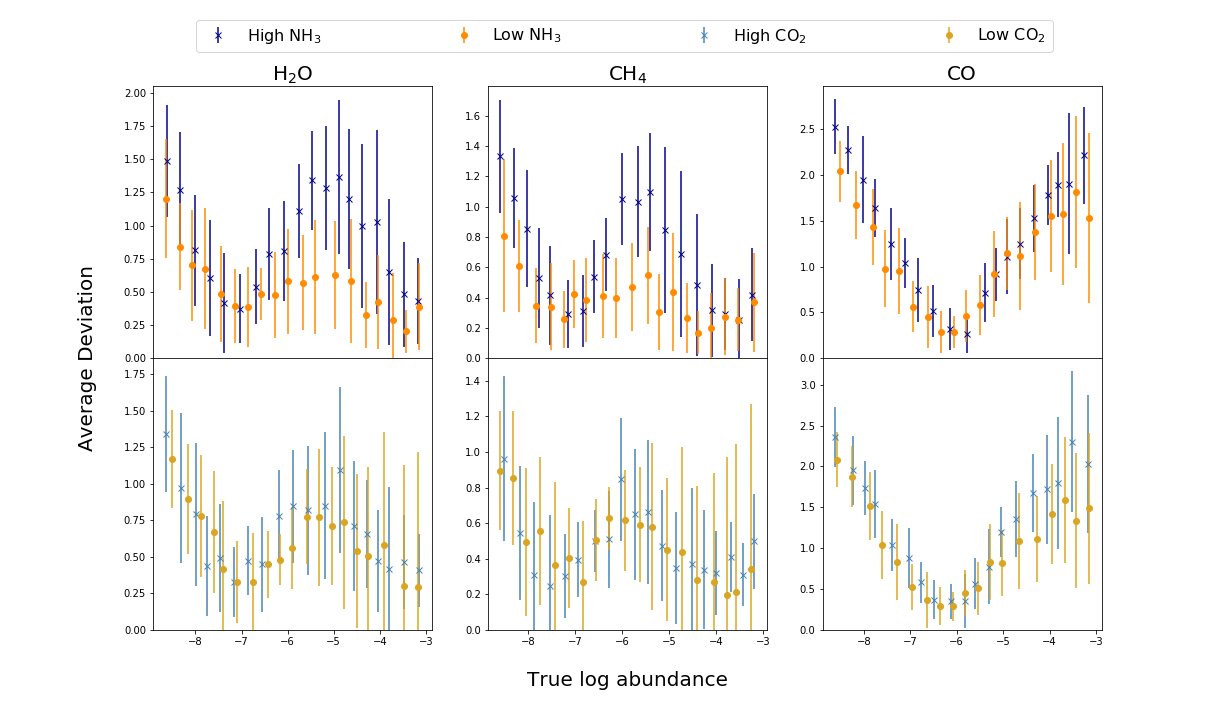}
    \caption{Visualization of bias \& variance for H$_2$O, CH$_4$ and CO at high and low NH$_3$ (top) and CO2 (bottom) (log-) abundance. H$_2$O and CH$_4$ prediction improved at low NH$_3$ abundance level. Please note that the model's variance also includes contributions from the irreducible noise.}
    \label{fig:BVPlot_NH3}
\end{figure*}
Our example above has investigated the average deviation and spread of the prediction at each prediction level. This quantification and visualization of bias and variance can be further utilised to help us determine the optimal model complexity for a given predictive task (see Appendix \ref{sec:complexity} for a detailed discussion).

\subsection{Interactions between AMPs}
\label{sec:interaction}
Next, we inspect whether the model aligns with our physical intuition on the problem. More specifically, we ask ``Does it perform worse when we expect it to?" One way to investigate this is to measure how its predictions on the $j$-th AMP $\hat{y_j}$ vary conditioned on the true value of another AMP $y_k$, $k \neq j$ having a `low' or a `high' value. For the purposes of this visualisation, we focused on gases and defined any log abundance in the lowest quartile (i.e. the lowest 25\% of the values) of the population as `low' and any log abundance in the highest quartile (i.e. the highest 25\% of the values) as `high'. 

Figure \ref{fig:BVPlot_NH3} shows how the (binned) predictions of H$_2$O, CH$_4$ and CO change under high ($> -4$) and low ($< -7$) abundances of NH$_3$ and CO$_2$. The binning procedure is similar to the one described Section \ref{sec:BVPlot}, but the bin size is reduced to 30 samples. 

We can observe that a high or low abundance of NH$_3$, gives rise to a distinctive contrast in the quality of predictions for most molecules. In particular, the quality of the predictions of CH$_4$ is highly affected by the abundance level of NH$_3$. This observation aligns with our expectations. As ammonia's absorption feature spans from 2 - 4 $\mu$m, it can partially or fully cover any other absorption features within that range at high abundances, reducing the model's ability to accurately predict the abundance of molecules such as CH$_4$, and vice versa. The same issue, however, should not arise for H$_2$O, as the molecule possesses several broad band features outside 2 - 4 $\mu$m, i.e. a well-trained network should be able to rely on information available outside this range to predict water abundance, which means there shouldn't be a dramatic improvement when NH$_3$ is low. This somewhat unexpected improvement in performance hints to the mechanism behind the model's prediction, this mechanism is further discussed in Section \ref{sec:resultsII}

On the other hand, the model's performance on CH$_4$ does not change as much under different levels of CO$_2$. This is also an expected outcome, as CO$_2$'s absorption feature lies in 5 - 6 $\mu$m, thus distinct features in CH$_4$ are less likely to be masked by changes in CO$_2$'s features \citep{sharp2007}.

It is possible to construct similar plots for other AMPs beyond NH$_3$ and CO$_2$. The purpose of this work is to demonstrate general evaluation tools that shed light into the inner workings of machine learning models. To avoid overly-emphasising our analysis on our particular (dataset, model) combination, we shall forgo an exhaustive discussion on all combinations of AMPs interactions\footnote{Interested readers can inspect the full results in Figures \ref{fig:h2o_bvp} - \ref{fig:cloud_bvp}, the results shown here are produced using our 1D-CNN model.}.

\begin{figure*}
    \centering
    \includegraphics[width=\textwidth]{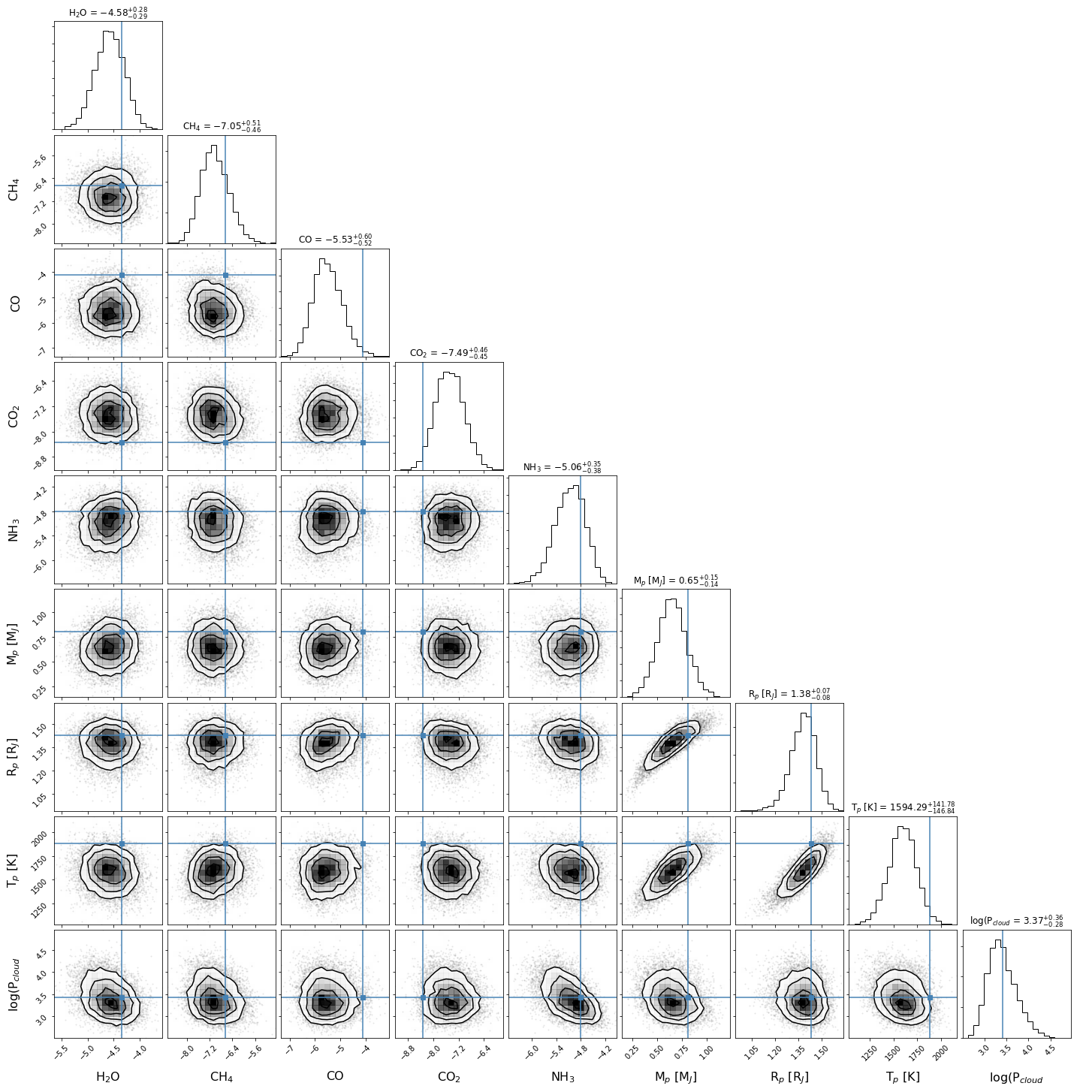}
    \caption{Posterior distribution from a test set spectrum. The ground truth is indicated by the blue line. The model tends to perform well on abundant trace gases with the exception of CO. It is also able to capture some correlation between AMPs.}
    \label{fig:posterior}
\end{figure*}

Another well-known way to visualise the covariance between different AMPs is to visualise the learnt posterior distribution. We sample the parameter space of the model by varying the same input spectrum according to its uncertainty. Figure \ref{fig:posterior} shows an example posterior distribution produced using a spectrum from the test set. The model managed to predict within 2-$\sigma$ of the ground truth values (indicated by the blue line) when the log abundances of the gases are higher than -6, which is expected from our analysis in previous sections.  We can see that the model is able to capture some of the correlations such as M$_p$ vs R$_p$, R$_p$ vs T$_p$, which are among some of the worst performing AMPs. On the other hand, the model failed to capture other well-known correlations such as H$_2$O vs R$_p$ and H$_2$O vs clouds. This analysis is thus highlighting a shortcoming of this model.

Upon discovering ways in which a model's behaviour is poor (either in terms of predictive performance, or in terms of capturing aspects of the underlying physics) we can take further measures to improve it.
In our analysis above, we noticed that certain known correlations among the targets (AMPs) were not captured by the model. There are ways to explicitly introduce domain knowledge like this into the architecture of the neural network, e.g. by parameter sharing across targets as discussed in \cite{reyes2019performing}. As this work is focused on analysing models and diagnosing problems, applying such methods to this setting is reserved for future work.

\subsection{Credibility of predictions}
\label{sseq:credibility}
\begin{figure}
    \centering
    \includegraphics[width=\columnwidth]{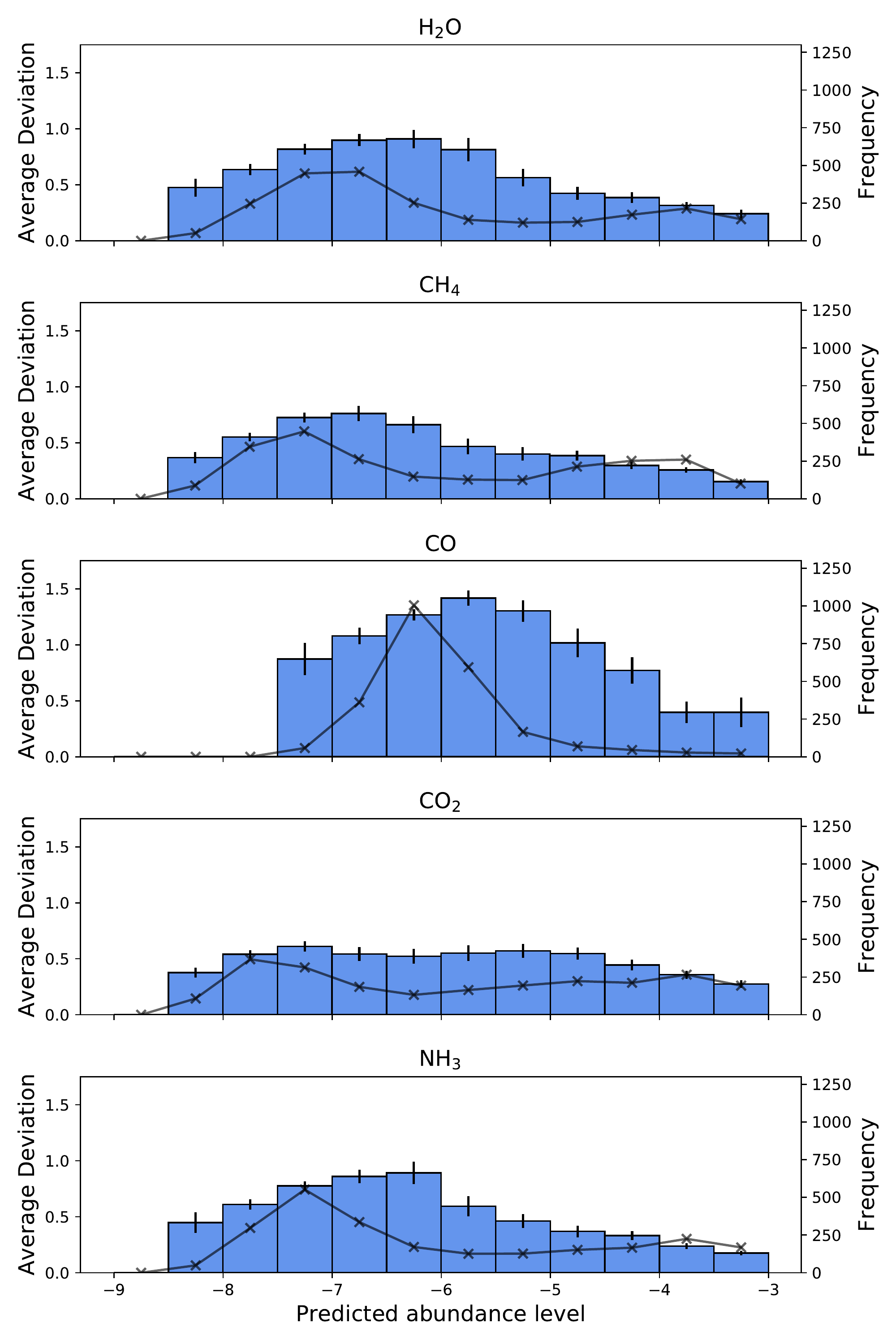}
    \caption{Average deviation for different molecules at different predicted abundance. The error bars on each bar represent 95 \% confidence intervals around the mean. The black curve represents the total number of predictions made by the model (frequency) within each bin. }
    \label{fig:deivation}
\end{figure}

So far we have looked at how the quality of a model's predictions varies across different ground truth values. During the training phase, this is useful for identifying problems with our models; targets whose prediction is problematic, areas of particularly high bias or variance, interactions among AMPs suggestive of degeneracies, among others.

However, in real life we rarely have access to the ground truth, thus a more practical question would be: ``What is the expected deviation of a given prediction from the ground truth?". In other words, we would like to know how credible a given prediction by our model is.

Our analysis in Sections \ref{sec:BVPlot} and \ref{sec:interaction} has provided us with some qualitative intuition regarding the credibility of the model's predictions. For example, the model's prediction is more reliable at higher molecular abundances and when there is less interference from other molecules. To obtain a more quantitative measure, we follow similar approach to Section \ref{sec:BVPlot} and compute the average deviation, at different prediction levels for each AMP (Figure \ref{fig:deivation}). The error bars on each bin represent 95\% confidence intervals. Instead of binning with equal frequency like Section \ref{sec:BVPlot}, we performed equal-width binning and thus each bin will have different number of data points. The black line shows the number of data points per bin. Bins with fewer than 20 data points are omitted.

The distribution of average deviations aligns with our discussion in Section \ref{sec:BVPlot}. For high predicted abundances the model starts off with low average deviation, and as the predicted abundance level goes down, the model struggles to predict well and begins to have higher average deviation. However, counter to our intuition, the average deviations do not increase monotonically and begin to decrease after a certain abundance level. This peak corresponds to the trough we saw in the figures of Section \ref{sec:BVPlot}.  This provides us with clues about of the model's loss minimisation strategy. The model is restricting its output to a limited range of values at low abundance levels, centred around some average value. This can be evidenced by the distribution of counts (black line) being centred at some value in the low abundance region and few or zero counts at the lowest abundance level (log(X$_{gas}$) = -9). 

Another important insight that can be drawn from Figure \ref{fig:deivation} is the trustworthiness of the prediction varies across abundance levels. We propose a method to qualitatively assess the credibility limit of each gas - the limit at which predictions remain meaningful to the model's user. First, we compute the probability $P$ that the model's prediction ($\hat{y_j}$) does not deviate more than a positive real value $\epsilon$ from the ground truth $y_j$. We then require that $P$ be at least $1-\delta$ to consider the prediction credible. A detailed discussion on the method is included in Appendix \ref{app:credibility}. In Figure \ref{fig:cred_limit} we demonstrate an example where we have chosen $\epsilon$ = 0.5, and defined a credibility threshold $\delta$ = 0.3, so that any prediction level with probability P $\geq 1-\delta$ is credible. We can then define the lowest predicted abundance level that satisfies this as the credibility limit of that gas.   

\begin{figure}
    \centering
    \includegraphics[width=\columnwidth]{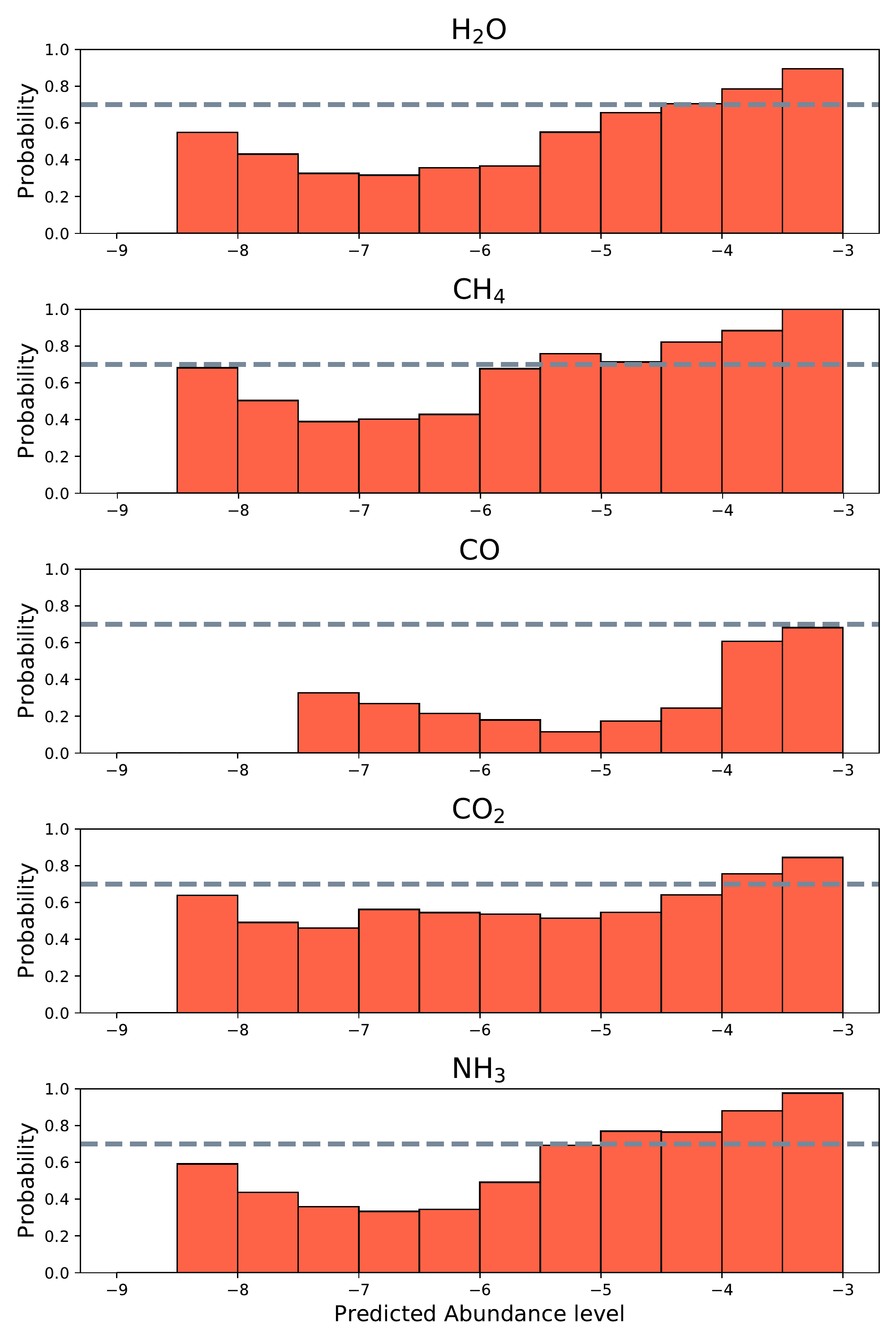}
    \caption{Probability of finding a prediction within $\epsilon$ = 0.5 to the ground truth for each molecule. The grey dotted line represents the probability threshold at 0.7 ($1-\delta$). Any bin with P $>$ 0.7 is considered credible in this case. }
    \label{fig:cred_limit}
\end{figure}


This limit is specifc to the chosen $\delta$ and $\epsilon$, as well as the trained model.   Table \ref{tab:upper_bound} summarises the estimated limit for each molecule. Although this approach is useful, we should still be mindful of its limitations. Note that for several gases the probability that the prediction error will not deviate from the acceptable region seems to increase at the lower end of the log abundance. However, this increase does not necessarily imply a higher predictive power at low abundances. In the aforementioned cases, this apparent increase in predictive power can be most likely attributed to the small number of instances that fall within these bins, as evidenced by Figure \ref{fig:deivation}. It is thus a small sample effect.

Similar work by \cite{alfnoor} determined the detection limit from a retrieval perspective. In their work they have generated 164 planets using the same setup as the one presented in Section \ref{sec:gen_data} and determined the lowest abundance at which they can constrain the molecular abundance within 1 order of magnitude of error. Our limit here addresses the trustworthiness of the neural network, which can not be directly compared with results from retrievals. Despite the differences, both studies suggest that Tier-2 Ariel spectra are capable of allowing for the consistent detection of some molecules in abundances as low as log(X) = -5.8. While it is possible that given a different architecture, the credibility limit could be improved, we would like to re-iterate that our goal is not to compete against retrieval frameworks, and thus we will leave this for future work.

Interested readers could refer to Figure \ref{fig:all_deviation} for the distribution of average deviation of all AMPs. Credibility limits for non-gaseous AMPs are less straightforward, and are more influenced by the training set. As the training set is not uniformly distributed w.r.t. these AMPs, trained models will have a tendency to focus on the regions containing a higher density of examples, biasing the predictions. Thus, the derived limit first and foremost would depend on biases of the training set and for this reason we chose to omit it here to avoid over-interpretation. 

\begin{table}[]
\centering
\begin{tabular}{l|l|l}
Molecules & Credibility Limit (log$_{10}$) & Detection Limit \\ \hline
H$_2$O & -4.3 & -6.5\\
CH$_4$ & -5.8 & -7 \\
CO & N/A & -5.5 \\
CO$_2$ & -3.8 & -7 \\
NH$_3$ & -5.3 & -6.5 
\end{tabular}
\caption{Credibility limit for each molecule at $\delta = 0.3,\epsilon=0.5$. This limit is derived using the lowest credible (log-) abundance level, following the credibility definition in Appendix \ref{app:credibility}. The detection limit is reproduced from \cite{alfnoor} as a comparison to retrieval methods}
\label{tab:upper_bound}
\end{table}
\section{Sensitivity Analysis for Model Interpretation}
\subsection{Method}
\label{sec:sensi_method}
Given any trained predictive model $\mathcal{M}$ (e.g. a neural network) that takes an input $\mathbf{x}$ and outputs the corresponding prediction $\hat{\mathbf{y}}$, a perturbation-based sensitivity test can be performed to assess the change in the prediction $\hat{y}$ when a set of features (transit depth, in our case) $x_i$ (consecutive or not) is perturbed.

 This approach assesses quantitatively how $\mathbf{\hat{y}}$ varies as a set of $x_i$ (transit depths) vary. The intuition is that perturbations in the regions of the input containing higher information about the target will yield larger deviations in the output of the model.
 
Below we outline a general procedure for such a sensitivity analysis:

\begin{enumerate}
  \item Produce a reference prediction $\mathbf{\hat{y}}_r$ on an unperturbed input $\mathbf{x}_r$. 
  \item Perturb the input $\mathbf{x}_p$.
  \item Predict $\mathbf{\hat{y}}_p$ on the perturbed input $\mathbf{x}_p$. 
  \item Compare $\mathbf{\hat{y}}_p$ and $\mathbf{\hat{y}}_r$.
  \item Repeat step 2 - 4 for different sets of features. 
\end{enumerate}

The form of perturbation depends on the context of the problem. \cite{occlustion} demonstrated the idea on models performing image classification. They perturbed the input image by setting a region to zero pixel value, and produced a heat-map of sensitivity by covering each region systematically. 
In this investigation we adapted this procedure to our multi-target regression problem. Instead of setting $x_i$ to zero like \cite{occlustion}, we applied the perturbation by sampling each wavelength bin $x_i$ according to its respective error bars (i.e. from a Gaussian centred at its unperturbed value and with standard deviation $\sigma_i$). There are three main reasons for this choice: 1. \emph{Physical plausibility.} Setting a window of the spectrum to zero would render it nonphysical, as a transit depth of 0 would mean $R_p = 0.$ 2. \emph{Statistical plausibility.} Neural networks excel at interpolation but not extrapolation. Perturbing the input spectrum within its error bars would still result in valid input (i.e. a sample from the actual data distribution) for the Neural Network. 3. \emph{Instrument plausibility.} The result of the test under these conditions provides realistic measurement of the relative sensitivity of each wavelength bin for the purposes of determining each of the parameters to be retrieved, in the context of Ariel Deep survey  specifications (Tier 2 spectra). This also means any derived result will be specific to the instrument and observing strategy. 

At each iteration we select a random number of $x_i$ and apply the perturbation by scattering these points according to $\mathcal{N}(x_i,\sigma_i^2)$. For computational efficiency, at each iteration the number of $x_i$ is chosen from 27 (half of the total number of wavelength bins) down to 2 (parts of a feature). The intention is to account for the influence from both broad and narrow features, as well as the inter-dependencies between different wavelength bins.  We repeat the above procedure 1000 times with 300 spectra randomly chosen from the test set and calculate the average mean squared difference per AMP (i.e. parameter to be retrieved) between $\mathbf{\hat{y}}_p$ and $\mathbf{\hat{y}}_r$ for each wavelength bin.  The result is a sensitivity map of the model's output  for each AMP, w.r.t. each feature. A detailed implementation of the test is discussed in Appendix \ref{app:sensi_imple}.

The sensitivity map is a tool for us to visualise what factors drive the model's predictions. As such, it allows us to investigate whether the model aligns with our physical intuition. This can also shed light to potential biases of the model or the training data. Finally, it can even aid us in identifying potentially undiscovered relationships among features.  

\subsection{Sensitivity map} 
\label{sec:resultsII}
\begin{figure*}
    \centering
    \includegraphics[width=\textwidth]{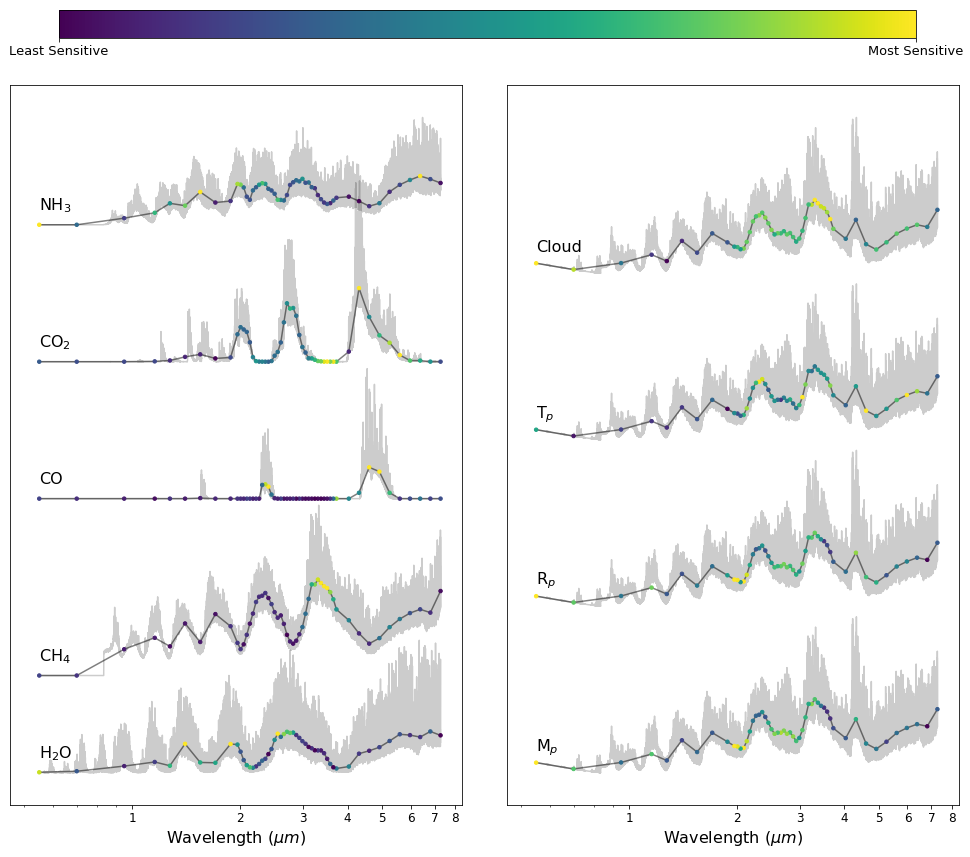}
    \caption{Sensitivity map for each AMP. Each wavelength bin is colour-coded to reflect the relative sensitivity level among bins, with yellow being the most sensitive and black being the least. Each spectrum shows the corresponding molecule’s characteristic absorption features in Ariel Tier-2 resolution.}
    \label{fig:sensi_map}
\end{figure*}

We applied the general procedure outlined in Section \ref{sec:sensi_method} and produced sensitivity maps for the different atmospheric parameters of interest, using as the model $\mathcal{M}$ the 1D-CNN we trained in Section \ref{sec:training}. We shall explicitly ignore regions where the model's prediction is uncertain and restrict the sensitivity analysis to cases with log(X$_{gas}$) $>$ -5 \footnote{There are different limits for some gases, but for simplicity we chose a conservative value.}. We will first investigate whether the model's predictions align with our physical intuitions regarding inferring AMPs from spectra. Figure \ref{fig:sensi_map} summarises the results of the sensitivity analysis for all the AMPs. The left sub-figure summarises the molecular species and the right sub-figure summarises the planetary parameters and clouds. 

\subsubsection{Sensitivity Map for active molecules.}
Each spectrum on the left sub-figure displays the corresponding molecule's characteristic absorption features in Ariel Tier-2 spectra and each bin in the spectrum is colour-coded to reflect the relative sensitivity of the model's prediction of the corresponding molecular abundance due to changes in the value of said bin. We normalised each spectrum according to its respective minimum and maximum.

We can see that many of the highlighted regions correspond to the major absorption features of the molecules. This alignment is  evidence that the neural network is recognising individual molecular features and basing its predictions of the corresponding molecular abundance on the peaks and troughs of these absorbing regions. Even in the case of CO, whose abundance the model generally fails to accurately predict, it nonetheless manages to highlight the absorption bands of the molecule as the most important region for predicting it.

So far, we see that the predictions of the neural network are based on factors that agree with our physical intuition. However, we can also see that for some molecules, not all peaks are highlighted by the model, i.e. for  H$_2$O only the peaks at 2-3$ \mu$m are highlighted. The model is tasked to jointly predict all quantities of interest. As a result, it is compromising performance across individual molecular species to identify the optimal features to predict them jointly. 

Sensitivity maps like these are useful for improving the transparency and thus, our confidence in the predictions of the model. On the other hand, they also give us an indication of where most of the information is coming from for the model in question. Here we only present maps for log(X$_{gas}) > -5$. It is possible that in the face of different combinations of abundances, the sensitivity map will change accordingly. As the purpose of this study is to explain the methodology, the discussion of sensitivity maps at different abundances will be left for future work. 

\subsubsection{Sensitivity Map for M$_p$,R$_p$, T$_p$ and clouds}
For AMPs other than gaseous species, their corresponding sensitivity maps are summarised on the right side of Figure \ref{fig:sensi_map}. We used the same, randomly sampled spectrum to investigate their sensitivity to each wavelength bin. Below we provide our observations and offer an interpretation of these maps. 

M$_p$, R$_p$, T$_p$ and clouds are interconnected via the computation of scale height, $H_{sc} = \frac{k_b T R_p^2}{\mu G M_p}$. The photometric points at the blue end of the spectrum are `calibration' points as they are the lowest points of the spectrum, these points help to provide an estimate for R$_p$ but are often masked in the presence of clouds. In the absence of M$_p$ from external sources, the model examined here attempts to derive it from the spectrum, which again is correlated with R$_p$, and clouds, as described in \cite{changeat2019}, and can be visualised via the similarities between their respective sensitivity maps.

Temperature on the other hand, is highlighted in three distinct regions, photometric waveband, ~3$\mu$m and ~5 $\mu$m regions. The model appears to be relying on the most probable highest features in the spectrum, combined with the photometric points, to derive the scale height via the features' size \footnote{The size of the features is determined by a quasi-linear function of H$_{sc}$.} and subsequently, the temperature. However, the aforementioned degeneracy between M$_p$, R$_p$, and clouds means that temperature is not accurately determined, as can be seen from Figure \ref{fig:BVPlot}. 

\subsection{Choice of Network}
We should keep in mind that sensitivity maps like the one shown in Figure \ref{fig:sensi_map} are model-specific, i.e. different models can have different sensitivity maps.
The sensitivity analysis described here does not directly measure the information in the features that is relevant for predicting the AMPs, but rather, it captures the degree to which a given model uses said information to predict the AMPs.

Figures \ref{fig:mlp_sensi} \& \ref{fig:lstm_sensi} show the sensitivity maps for the networks of the other two DNN architectures we trained as outlined in Section \ref{sec:training}. Interestingly,  
we find that all three models were able to highlight most of the peaks and troughs of the molecule’s characteristic features. This is evidence\footnote{Naturally, the class of models explored here is constrained to DNNs, so we cannot make any strong model-independent claims. Yet in this case, all three models agree with one another and with physical intuition.} that these regions of the spectra are indeed important features for determining their corresponding molecular abundances. 

There are also notable differences in the sensitivity maps of each model, in particular the maps obtained for non-molecular AMPs. For example, the MLP tends to focus on 4-5 $\mu$m to derive quantities such as M$_p$, R$_p$. T$_p$ and cloud top pressure, while the LSTM and the 1D-CNN tend to also focus on 2-3 $\mu$m features. Despite any differences in sensitivity across different models, non trace-gas AMPs exhibit high sensitivity to the same regions for a given model, highlighting the degeneracy between these AMPs.




\section{Conclusion} 
\label{sec:conclusion}
In the context of exoplanet atmospheric retrievals using simulated data from Ariel, we investigated the use of three different types of DNN architectures (MLP, CNN, LSTM) for inferring atmospheric
model parameters from exoplanet spectra. We presented a suite of methodologies for analysing the performance of any regression model, identifying its main source of error by leveraging the concepts of bias and variance, and quantifying the credibility of its predictions. Applying these evaluation methodologies to the three DNN models we trained, we found that they all behaved similarly for this dataset, and that they are capable of reliably determining molecular abundances down to as low as 10$^{-5.8}$.

We also introduced a perturbation based sensitivity analysis which allows us to assess the relative importance of each feature (wavelength bin) in predicting each target (atmospheric parameter), for a given trained predictive model. Our analysis confirmed that the predictions of the DNN models we constructed largely align with our physical intuition with respect to each atmospheric parameter's spectral signature,  our understanding of Ariel's instrument specification and Ariel Deep survey observational strategy. 

The evaluation and interpretability methods presented in this paper are applicable to any predictive model learned from data and only require access to the model's predictions and the training data. These tools allow us to analyse the predictions of a model, identify potential biases in the model itself or the data, understand the factors driving the model's predictions and investigate whether these agree with our current knowledge of the underlying physics, whether the model's predictions are `right for the wrong reasons' or whether it can provide us with new theoretical insights. Ultimately, they can make predictive models more transparent and thus more easy to adopt by domain experts.
\newline
\newline

\textbf{Software:} ArielRad: \citep{mugnai_Arielrad}, TauREx3 \citep{Taurex3},  h5py \citep{hdf5_collette},  Matplotlib \citep{Hunter_matplotlib}, Pandas \citep{mckinney_pandas}, Numpy \citep{oliphant_numpy}, Keras \citep{chollet2015keras}, Tensorflow \citep{tensorflow}. 
\newline
\newline
{\large \textbf{Acknowledgements}}

We appreciate suggestions from the anonymous reviewer, which has improved the quality of the manuscript. This project has received funding from the European Research Council (ERC) under the European Union’s Horizon 2020 research and innovation programme (grant agreement No 758892, ExoAI) and the European Union’s Horizon 2020 COMPET programme (grant agreement No 776403, ExoplANETS A). Furthermore, we acknowledge funding by the Science and Technology Funding Council (STFC) grants: ST/K502406/1, ST/P000282/1, ST/P002153/1 and ST/S002634/1. We are grateful for the support of the NVIDIA Corporation through the NVIDIA GPU Grant program.

\appendix
\section{Implementation details}
\label{app:imple}
Table \ref{tab:archi} summarises the architecture details of the three types of neural networks we explored. The hyperparameters were selected after performing a grid search on the number hidden units per layer, the number of layers and the number of filters. In all cases, the models were trained for 100 epochs with an initial learning rate of 0.01 and a learning rate decay of $10^{-4}$ using the Adam optimizer. Any unspecified hyperparameters were set to default Keras/Tensorflow values. All the networks were  developed using the open source \textbf{Keras} (Version 2.3.1) \textbf{python} module \citep{chollet2015keras}, with \textbf{Tensorflow} (Version 2.4.1) as backend \citep{tensorflow} . 

Table \ref{tab:performance} shows the average performance of each architecture across 5 runs with identical hyperparameter setup, but different weight initialisation and under different training/validation splits. We also compare their complexity as measured by the number of weights to be learned. All 3 architectures yielded models with comparable predictive performances. However, we chose to present our main results using the CNN due to its lower complexity and subsequently faster training \& inference time.

\begin{table}[]

\begin{tabular}{llc|llc|lll}
\multicolumn{9}{c}{Neural Network Architecture} \\ \hline \hline
\multicolumn{3}{c|}{MLP} & \multicolumn{3}{c|}{CNN} & \multicolumn{3}{c}{LSTM} \\ \hline
Layer Type & Config. & Output & Layer Type & Config. & Output & Layer Type & Config. & Output \\
Input &  & (m,52) & Input &  & (m,52,1) & Input &  & (m,52,1) \\
FC-RELU & h=320 & (m,320) & Conv-BN-RELU & f=32,3x3,s=1 & (m,52,32) & LSTM & h=200 & (m,52,200) \\
FC-RELU & h=240 & (m,240) & Maxpooling & 2x2 & (m,26,32) & LSTM & h=200 & (m,200 ) \\
FC-RELU & h=160 & (m,160) & Conv-BN-RELU & f=64,3x3,s=1 & (m,26,64) & FC-LeakyRELU & h=16 & (m,16) \\
FC-RELU & h=80 & (m,80) & Maxpooling & 2x2 & (m,13,64) & Dropout & p=0.3 & (m,16) \\
Dropout & p=0.3 & (m,80) & Conv-BN-RELU & f=96,3x3 ,s=1 & (m,13,96) & FC-Linear & h=9 & (m,9) \\
FC-Linear & h=9 & (m,9) & Flatten &  & (m,1248) &  &  &  \\
 &  &  & FC-leakyRELU & h=128 & (m,128) &  &  &  \\
 &  &  & Dropout & p=0.3 & (m,128) &  &  &  \\
 &  & \multicolumn{1}{l|}{} & FC-Linear & h=9 & (m,9) &  &  & 
\end{tabular}
\caption{The 3 different neural network architectures examined in this study. `BN', `FC' and `Conv' denote Batch Normalisation, Fully Connected and 1D Convolutional layer, respectively. With `h',`f',`s' and `p' we denote the hidden layer size, the filter size, the stride and the dropout probability, respectively. }
\label{tab:archi}
\end{table}

\begin{table}[]
\centering
\begin{tabular}{ccc}
\multicolumn{3}{c}{Best Performance versus Model Complexity} \\ \hline \hline
\multicolumn{1}{c|}{Architecture} & Performance & \# of weights \\ \hline
\multicolumn{1}{c|}{MLP} & 0.28 $\pm$ 0.01 & 248,889 \\
\multicolumn{1}{c|}{1D-CNN} & 0.28 $\pm$ 0.01 & 186,665 \\
\multicolumn{1}{c|}{LSTM} & 0.29 $\pm$ 0.01 & 325,769
\end{tabular}

\caption{Average performance of the 3 different architectures across 5 runs vs. model complexity (as measured by number of parameters).}
\label{tab:performance}
\end{table}

\section{Model Complexity}
\label{sec:complexity}
Quantifying the bias and variance of a model is also useful in determining whether the model is underfitting or overfitting the data. An underfitting model lacks the complexity (capacity) to learn the underlying pattern in the training set. This results in high error in the training set. It also results in poor performance on he test set. In this scenario, the prediction error is dominated by high bias. On the other hand, an overfitting model is ``excessively complex'' for the task at hand. This complexity can result in fitting not only the underlying pattern of interest, but also the noise in the training data, leading to a good fit on the training set, but poor generalisation in the test set. In this scenario, the prediction error is dominated by high variance. The ideal model for the task is the one with just enough complexity to neither underfit nor overfit.

If we identify that the model is underfitting on a given task, then we should increase its complexity. In the case of DNNs, this can be achieved by e.g. increasing the number of layers or the number of hidden units per layer. If we detect that a model is overfitting, one solution is to decrease the complexity of the model. We can either draw models from a richer model family (e.g. in the case of DNNs choose an architecture with more hidden layers and/or hidden units per layer), or we can introduce some form of regularization (e.g. $L_1$-regularization, $L_2$-regularization batch normalization or dropout). Alternatively, we can use an ensemble of several predictors (e.g. combine the predictions of multiple DNNs). Finally, if such a thing is possible, we can increase the amount of training data.

\begin{figure}

    \centering
    \includegraphics[width=\columnwidth]{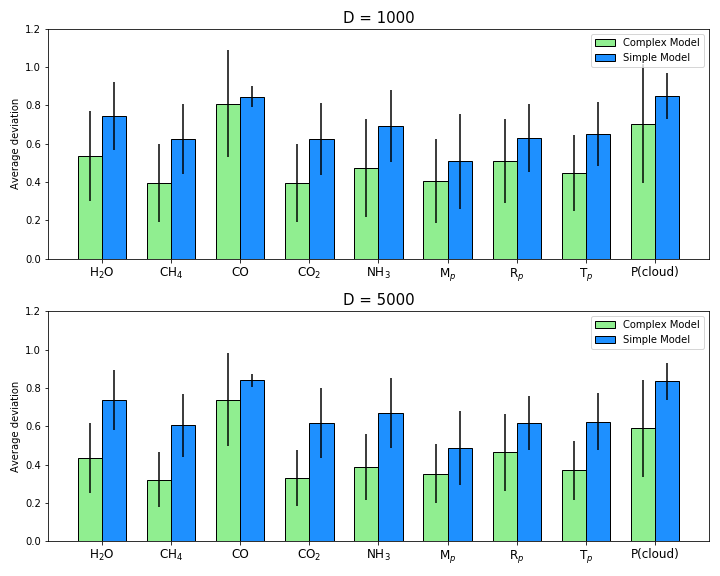}
    \caption{The performance of a `simple' and a `complex' model, on each AMPs, under a different amount of data points. The predictions of the `simple' model exhibit a higher average deviation from the true value (higher bias), whilst those of the `complex' model exhibit a higher variance. Training the two models on a larger training set allows the `complex' model to reduce its variance, but only marginally improves the predictions of the `simple model'.} 
    \label{fig:bvp_model}
\end{figure}

The example below demonstrates the effect of increasing the amount of training data available to a model that underfits the data (too low complexity, high bias) and a model that overfits the data (too high complexity, high variance). We first train two models, a `simple' and `complex' one. The `simple' model is consisted of two CNN layers with 8 filters each, and the `complex' model is composed of three CNN layers with 128 filters each. Both are trained  using 1000 training datapoints, repeating the data generation and training 20 times. The top part of Figure \ref{fig:bvp_model} shows the average deviation of the predictions of the two models for the different AMPs (a measure of the bias component of the prediction error), along with their standard deviation across the 20 runs (a proxy for the variance of the error). We then repeat the same experiment but this time we provide training 5000 datapoints to the two models. The bottom part of Figure \ref{fig:bvp_model} corresponds to the results when the two models are trained with an increased training sample size.

We can see that the `simple' model has a lower predictive performance than the `complex' one. Moreover, its performance only slightly improves when provided with more data. The results suggest that the error of the `simple' model is mainly due to bias and it cannot be decreased when trained on more data. The model has a limited capacity, smaller than the one required to model the dataset in this situation and thus underfits. On the other hand, the `complex' model exhibits a lower bias (as measured by the average deviation) than the `simple' one. However, its error is characterised by a notably larger variance (as approximated by the standard deviation) compared to the `simple' model. Finally, when provided with a larger training sample, the variance of its predictions decreases.


\section{Credibility Limit Calculations}
\label{app:credibility}
Consider random variables $Y$ and $\hat{Y}$ that represent the ground truth value of an AMP for a particular instance and a model's prediction on an AMP (i.e. logX$_{H_2O}$ abundance) respectively, we can define the credibility limit $L(\delta,\epsilon)$ for this AMP as the lowest possible predicted value $\hat{y}$ for which:
\begin{equation}
    P(Y=\hat{y} \pm \epsilon|\hat{Y}=\hat{y}) > 1- \delta
\end{equation}
where $P(Y=\hat{y} \pm \epsilon|\hat{Y}=\hat{y})$ is the probability of finding a model's prediction $\hat{y}$ within $\epsilon$ of the ground truth $y$. It can be computed for the entire set or per each abundance bin as:
\begin{equation}
    P=\frac{\#(Y\in |\hat{y}-\epsilon,\hat{y}+\epsilon| \cap \hat{Y}=\hat{y})}{\#(\hat{Y} = \hat{y})}
\end{equation}
where $\#(\cdot)$ denotes the number of datapoints that satisfy the conditions enclosed in the parentheses.

\section{Sensitivity Map Calculation: Implementation}
\label{app:sensi_imple}
This section provides a detailed description on how to compute the sensitivity map for a particular target. The notation follows that of the general procedure outlined in Section \ref{sec:sensi_method}. 

Given any trained predictive model $\mathcal{M}$ (e.g. a neural network) that takes an input spectrum $\mathbf{x}$ and outputs the corresponding prediction $\hat{\mathbf{y}}$, a perturbation-based sensitivity test can be performed to assess the change in the prediction $\hat{\mathbf{y}}$ when a set of features (transit depth, in our case) $x_i$ (consecutive or not) is perturbed. Algorithm \ref{seni_algo} and \ref{alg:perturb} demonstrates our procedure to compute the sensitivity map for a single spectrum on a single prediction.

Our implementation assumes no prior knowledge on the size and distribution of spectral features. Perturbation is applied to a randomly selected bin or group of bins at each iteration. Parameter $k$ controls the number of bins selected each time. Here we iterate through different values of $k$ to capture spectral features at different scales. To quantify the deviation of the perturbed prediction from the un-perturbed, we chose the Mean Squared Difference (MSE) due to its sensitivity to large differences, which helps to highlight sensitive regions.

The general procedure for generating sensitivity maps described in Section \ref{sec:sensi_method} is not tied to any specific distance measure, way of selecting the wavelength bins to be perturbed or perturbation method. The only requirement is that perturbed wavelength bin(s) be attributed a score based on the magnitude of the distance of the two predictions.

\begin{algorithm}
\caption{Sensitivity map calculation}\label{seni_algo}

    \KwData{un-perturbed spectrum $\boldsymbol{x_r}$, associated uncertainty vector $\boldsymbol{\sigma}$ and number of repetitions $N$}
    \KwResult{sensitivity score, \textbf{score}, of size ($|\boldsymbol{x_r}| \times |\boldsymbol{\hat{y}_r}|$) for each input feature (wavelength bin) $x_i$ and output variable AMP.}
\Begin{
// get prediction from an un-perturbed spectrum \;
$\boldsymbol{\hat{y}_r} \assign $\FuncCall{ModelPrediction}{$\boldsymbol{x_r}$} \;
// empty array to store results \;
$\textbf{delta} \assign $\FuncCall{Zeros}{$|\boldsymbol{x_r}|$, |$\boldsymbol{\hat{y}_r}$|} \;
$l \assign $ $|\boldsymbol{x_r}|$ \;

\For{$n \assign 0 $ \KwTo $N-1$}{
    $\boldsymbol{L} \assign $\FuncCall{Zeros}{$|\boldsymbol{x_r}|$, |$\boldsymbol{\hat{y}_r}$|} \;
    // Extract position index from an array \;
    $\textbf{index} \assign $\FuncCall{GetIndex}{$\boldsymbol{x_r}$} \;
    // shuffle index at random \;
    $\textbf{shuffled\_index} \assign $\FuncCall{Shuffle}{$\textbf{index}$} \;
    $k \assign l $ \;
    \While{$k \geq 2$}{
    $k \assign $ \FuncCall{Ceil}{$k/2$}  \;
    // randomly draw k positions without repetition\;
    $\textbf{pos} \assign $ \FuncCall{DrawWithoutRepetition}{$\textbf{shuffled\_index},k$} \;
    $\boldsymbol{x_p} \assign$ \FuncCall{Perturb}{$\boldsymbol{x_r},\boldsymbol{\sigma},\textbf{pos}$} \;
    $\boldsymbol{\hat{y}_p} \assign $ \FuncCall{ModelPrediction}{$\boldsymbol{x_p}$} \;

    // assign $\boldsymbol{\hat{y}_p}$ to $\boldsymbol{L}$ at position $\textbf{pos}$ \;
    $\boldsymbol{L}[\textbf{pos},\boldsymbol{:}] \assign \boldsymbol{\hat{y}_p}$ \;
    }
    // user defined distance metric \;
    $\textbf{diff} \assign$ \FuncCall{SquaredDifference}{$ \boldsymbol{\hat{y}_r}, \boldsymbol{L}$} \;
    $\textbf{delta} \assign \textbf{delta} + \textbf{diff}$ \;
    $n \assign n + 1 $ \;
    }
$\textbf{score} \assign \textbf{delta}/N$ \;
}

\end{algorithm}

\begin{algorithm}    
\label{alg:perturb}
    \caption{Perturbation method used in this investigation.}
    \underline{\textbf{Function} Perturb}$(\boldsymbol{x_r}, \boldsymbol{\sigma}, \textbf{pos})$\;
    \KwData{un-perturbed spectrum $\boldsymbol{x_r}$, associated uncertainty vector $\boldsymbol{\sigma}$ and set of positions in which to apply perturbation $\textbf{pos}$}
    \KwResult{perturbed spectrum $\boldsymbol{x_p}$}
\Begin{
$\boldsymbol{x_p} \assign$ \FuncCall{Copy}{$\boldsymbol{x_r} $} \;
$\textbf{sample} \assign $  \FuncCall{DrawFromGaussian}{$\boldsymbol{x_p}[\textbf{pos}], \boldsymbol{\sigma} [\textbf{pos}]$} \;
// replace values at position $\textbf{pos}$ with perturbed values \;
$\boldsymbol{x_p}[\textbf{pos}] \assign \textbf{sample} $ \;
}
\end{algorithm}

\begin{figure}
    \centering
    \includegraphics[width=\columnwidth]{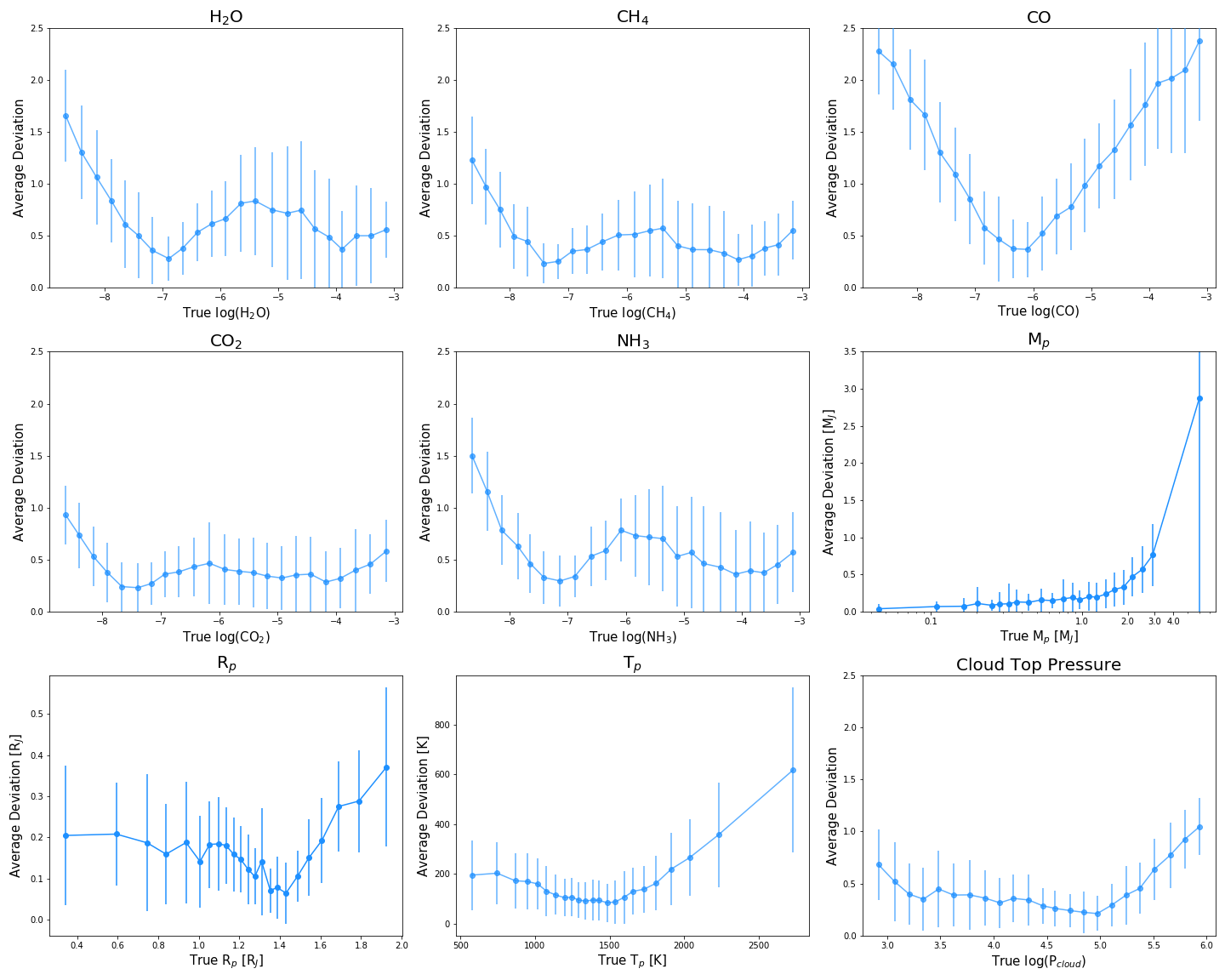}
    \caption{Visualization of bias \& variance for different AMPs produced by the MLP model in their actual units. Each point represents the average deviation at that level and its error-bar represents the 1-$\sigma$ spread of the prediction. The performance is mostly similar to the one derived from the CNN model. } 
    \label{fig:mlp_bvp}
\end{figure}

\begin{figure}
    \centering
    \includegraphics[width=\columnwidth]{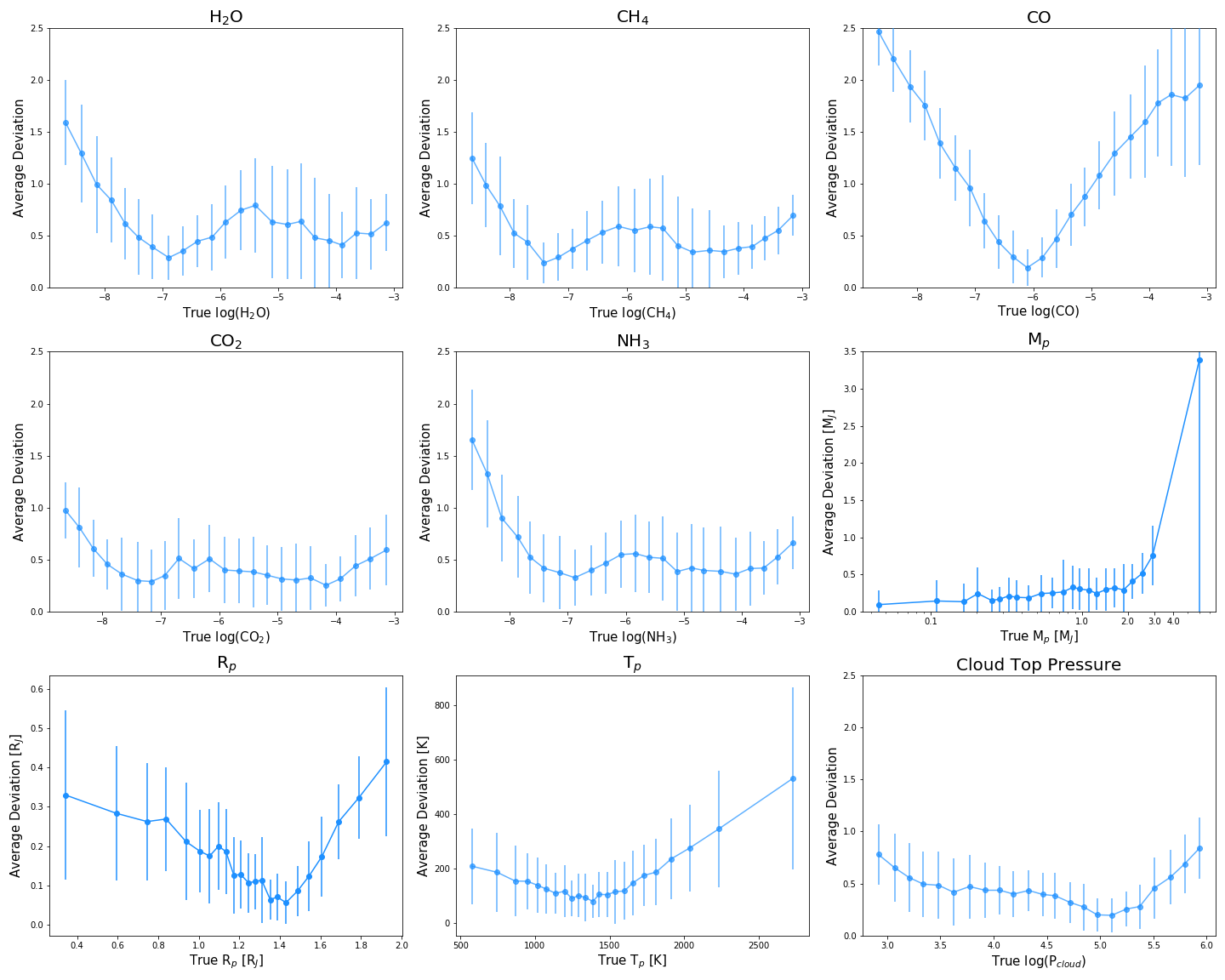}
    \caption{Visualization of bias \& variance for different AMPs produced from the LSTM model. Each point represents the average deviation at that level and its error-bar represents the 1-$\sigma$ spread of the prediction. The performance is mostly similar to the one derived from the CNN model.} 
    \label{fig:lstm_bvp}
\end{figure}

\begin{figure}
    \centering
    \includegraphics[width=\columnwidth]{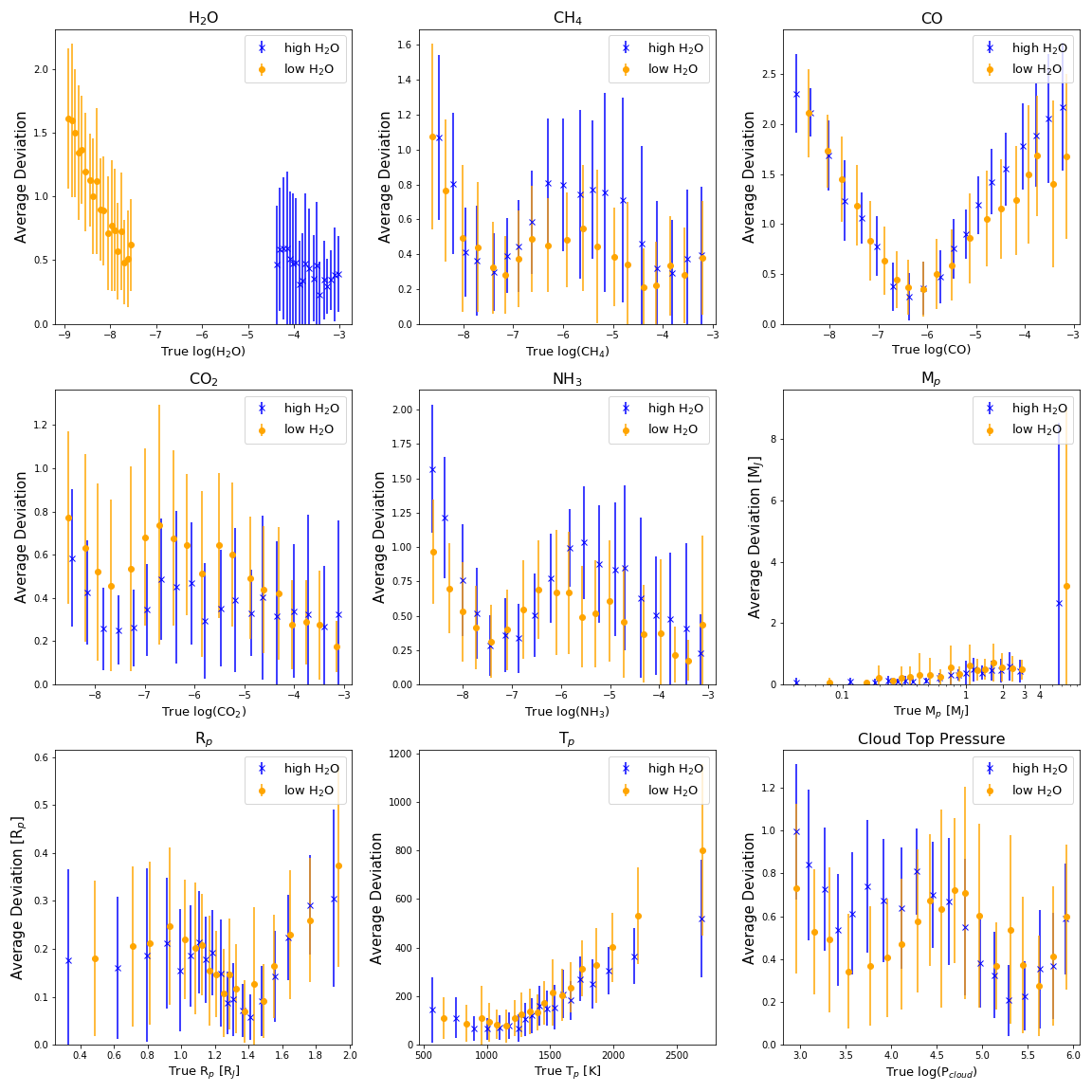}
    \caption{Visualization of bias \& variance for different AMPs at high and low H$_2$O (log-)abundance. Each point represents the average deviation at that level and its error-bar represents the 1-$\sigma$ spread of the prediction. This plot is generated using the CNN model.} 
    \label{fig:h2o_bvp}
\end{figure}

\begin{figure}
    \centering
    \includegraphics[width=\columnwidth]{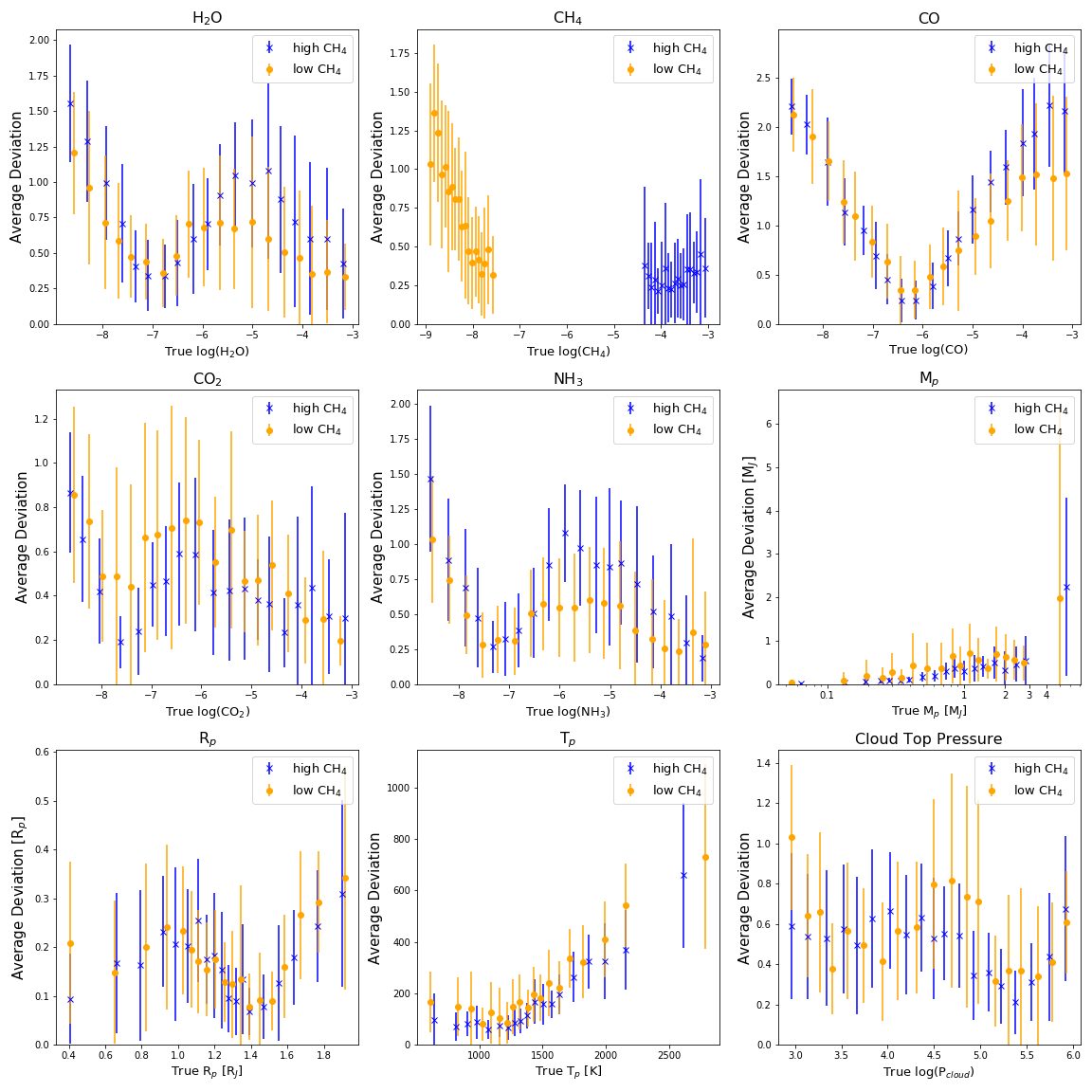}
    \caption{Visualization of bias \& variance for different AMPs at high and low CH$_4$ (log-)abundance. Each point represents the average deviation at that level and its error-bar represents the 1-$\sigma$ spread of the prediction. This plot is generated using the CNN model.} 
    \label{fig:ch4_bvp}
\end{figure}

\begin{figure}
    \centering
    \includegraphics[width=\columnwidth]{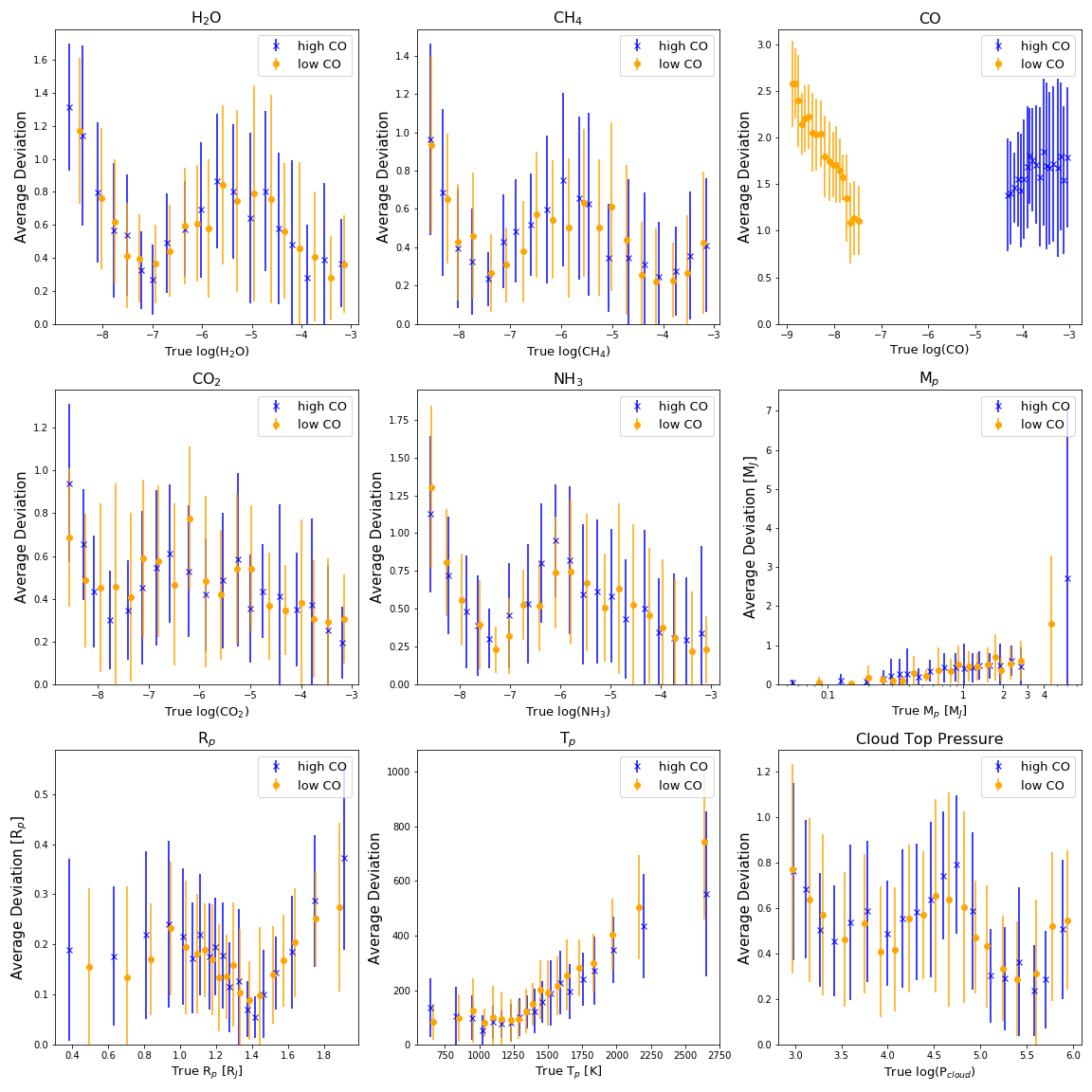}
    \caption{Visualization of bias \& variance for different AMPs at high and low CO abundance. Each point represents the average deviation at that level and its error-bar represents the 1-$\sigma$ spread of the prediction. This plot is generated using the CNN model.} 
    \label{fig:co_bvp}
\end{figure}

\begin{figure}
    \centering
    \includegraphics[width=\columnwidth]{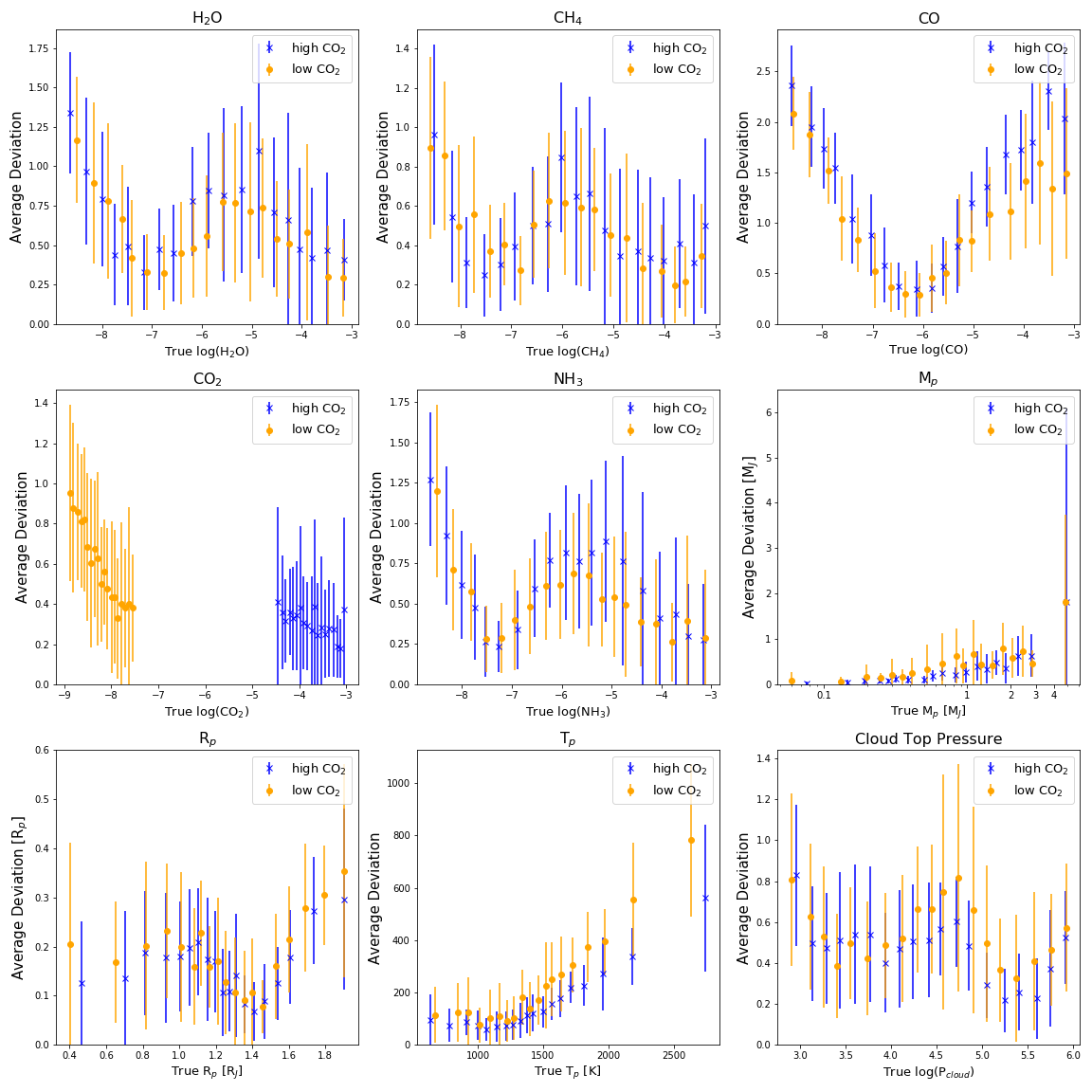}
    \caption{Visualization of bias \& variance for different AMPs at high and low CO$_2$ (log-)abundance. Each point represents the average deviation at that level and its error-bar represents the 1-$\sigma$ spread of the prediction. This plot is generated using the CNN model.} 
    \label{fig:co2_bvp}
\end{figure}

\begin{figure}
    \centering
    \includegraphics[width=\columnwidth]{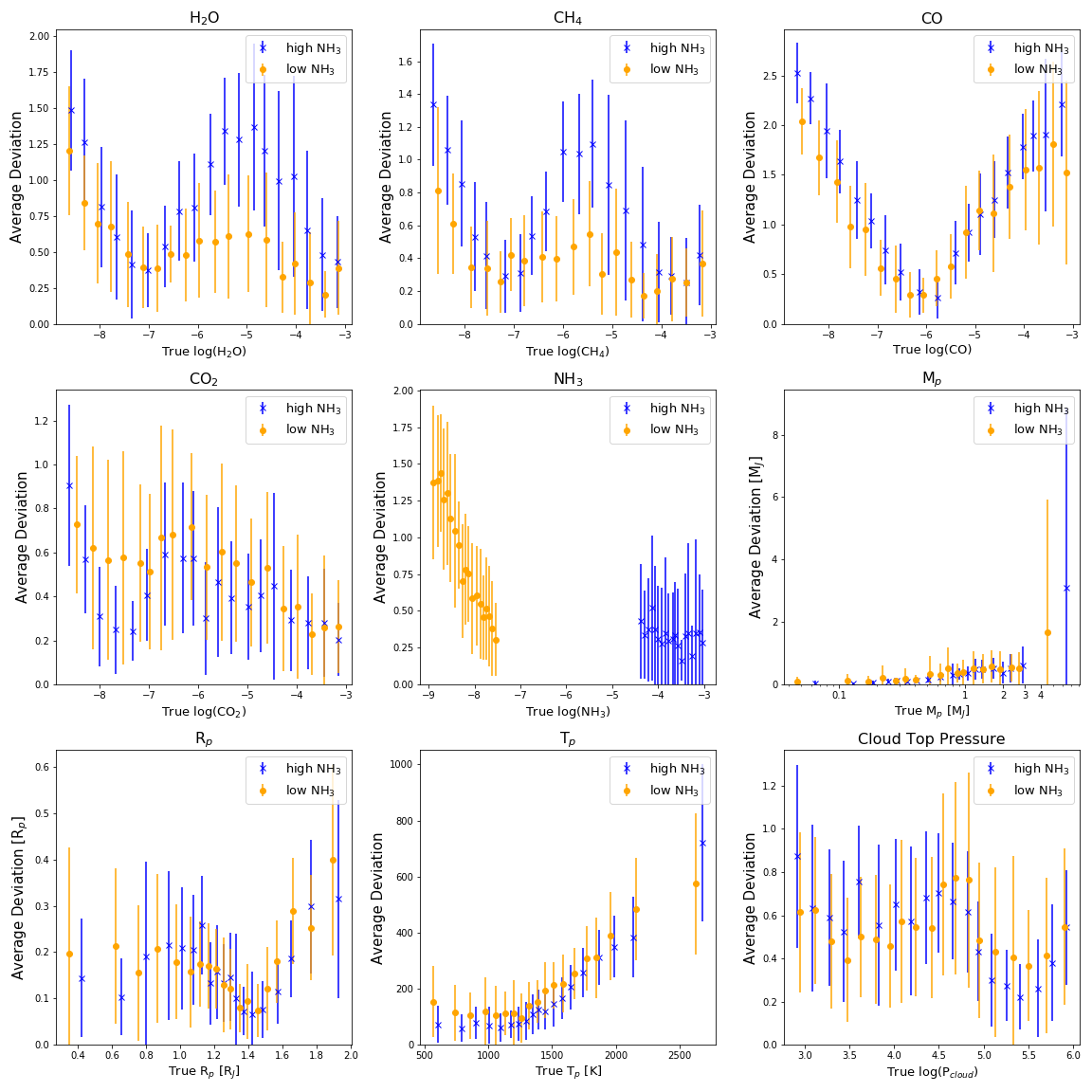}
    \caption{Visualization of bias \& variance for different AMPs at high and low NH$_3$ abundance. Each point represents the average deviation at that level and its error-bar represents the 1-$\sigma$ spread of the prediction. This plot is generated using the CNN model.} 
    \label{fig:nh3_bvp}
\end{figure}

\begin{figure}
    \centering
    \includegraphics[width=\columnwidth]{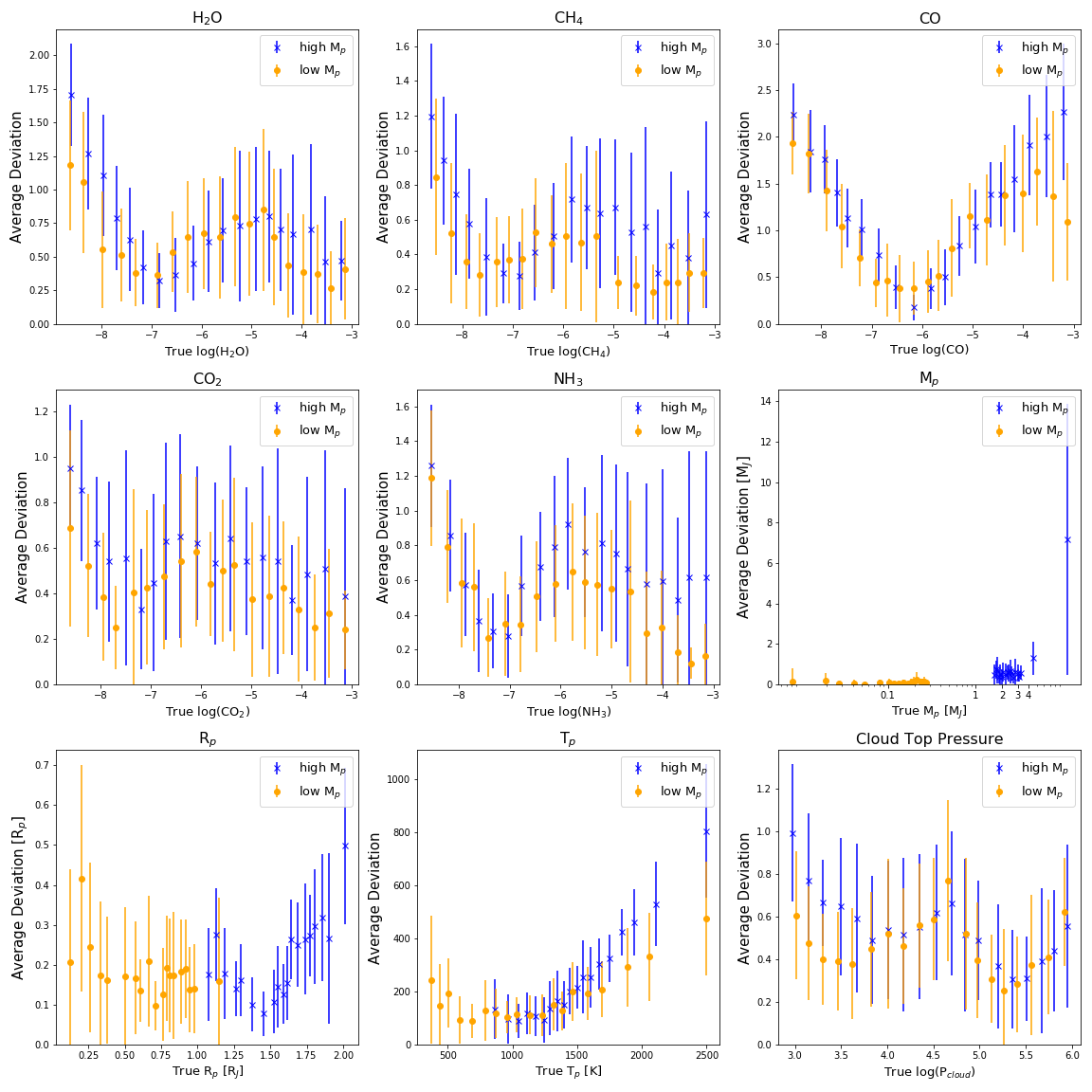}
    \caption{Visualization of bias \& variance for different AMPs at high and low M$_p$. Each point represents the average deviation at that level and its error-bar represents the 1-$\sigma$ spread of the prediction. This plot is generated using the CNN model.} 
    \label{fig:mp_bvp}
\end{figure}

\begin{figure}
    \centering
    \includegraphics[width=\columnwidth]{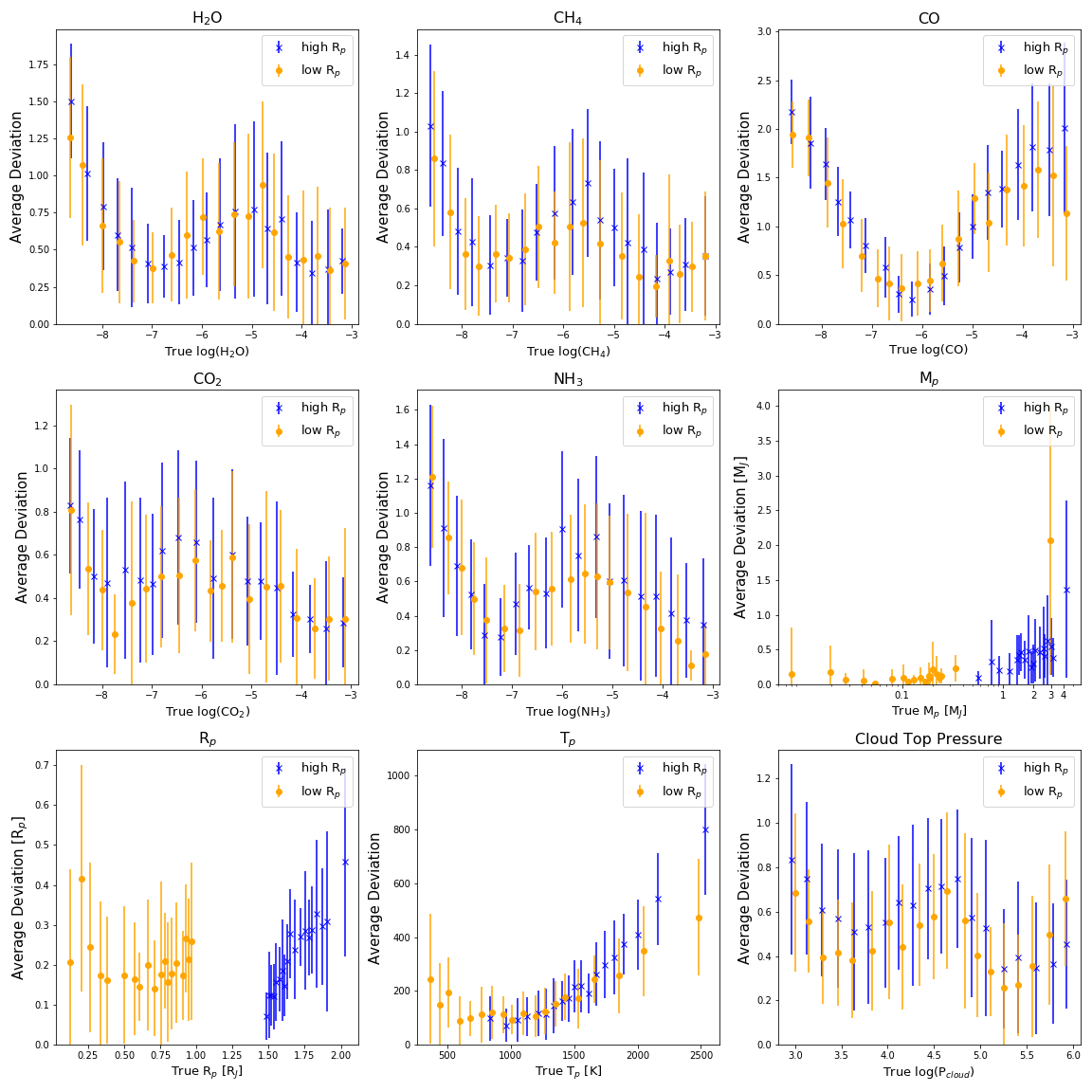}
    \caption{Visualization of bias \& variance for different AMPs at high and low R$_p$. Each point represents the average deviation at that level and its error-bar represents the 1-$\sigma$ spread of the prediction. This plot is generated using the CNN model.} 
    \label{fig:rp_bvp}
\end{figure}

\begin{figure}
    \centering
    \includegraphics[width=\columnwidth]{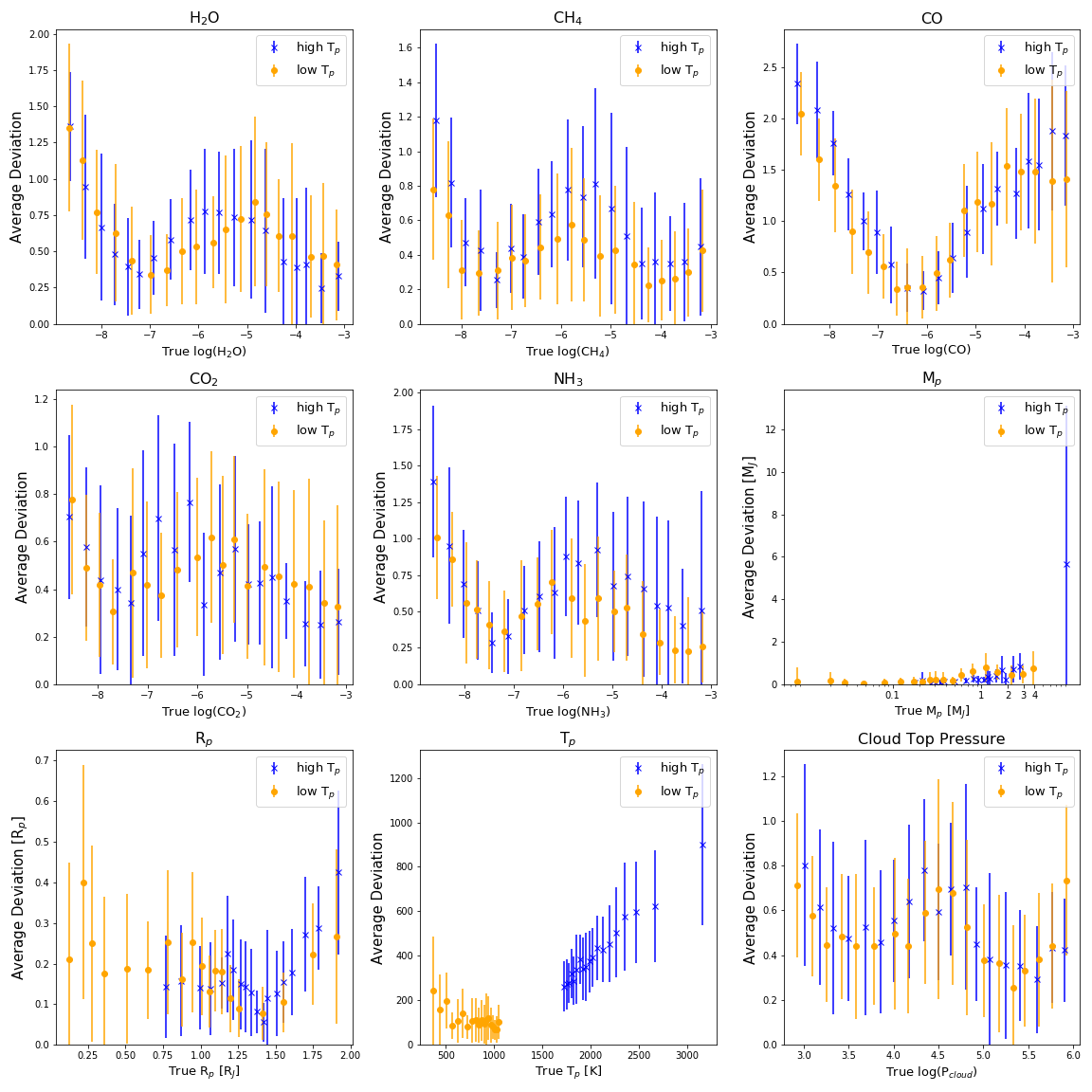}
    \caption{Visualization of bias \& variance for different AMPs at high and low T$_p$. Each point represents the average deviation at that level and its error-bar represents the 1-$\sigma$ spread of the prediction. This plot is generated using the CNN model.} 
    \label{fig:tp_bvp}
\end{figure}

\begin{figure}
    \centering
    \includegraphics[width=\columnwidth]{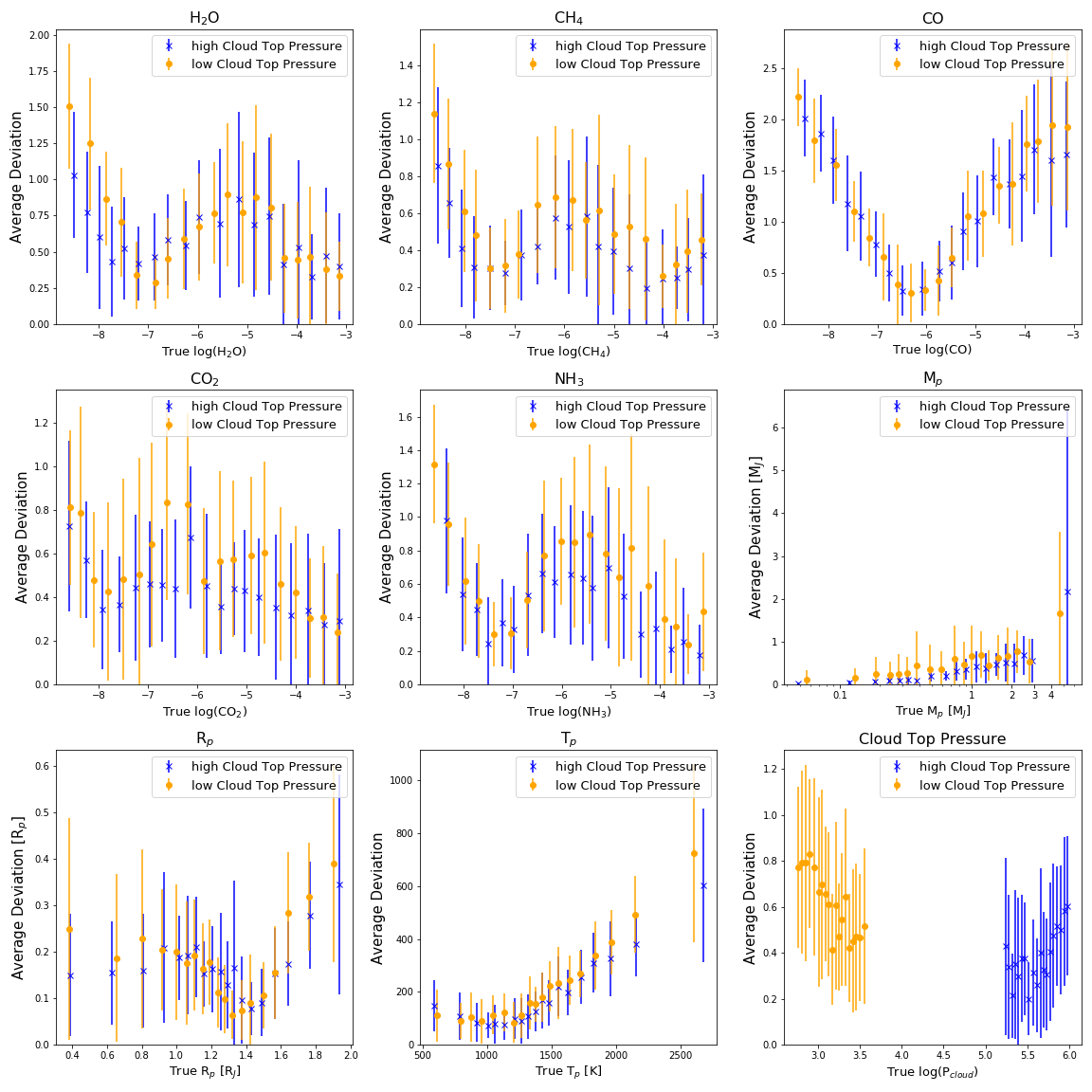}
    \caption{Visualization of bias \& variance for different AMPs at high and low Cloud Top Pressure. Each point represents the average deviation at that level and its error-bar represents the 1-$\sigma$ spread of the prediction} 
    \label{fig:cloud_bvp}
\end{figure}

\begin{figure}
    \centering
    \includegraphics[width=\columnwidth]{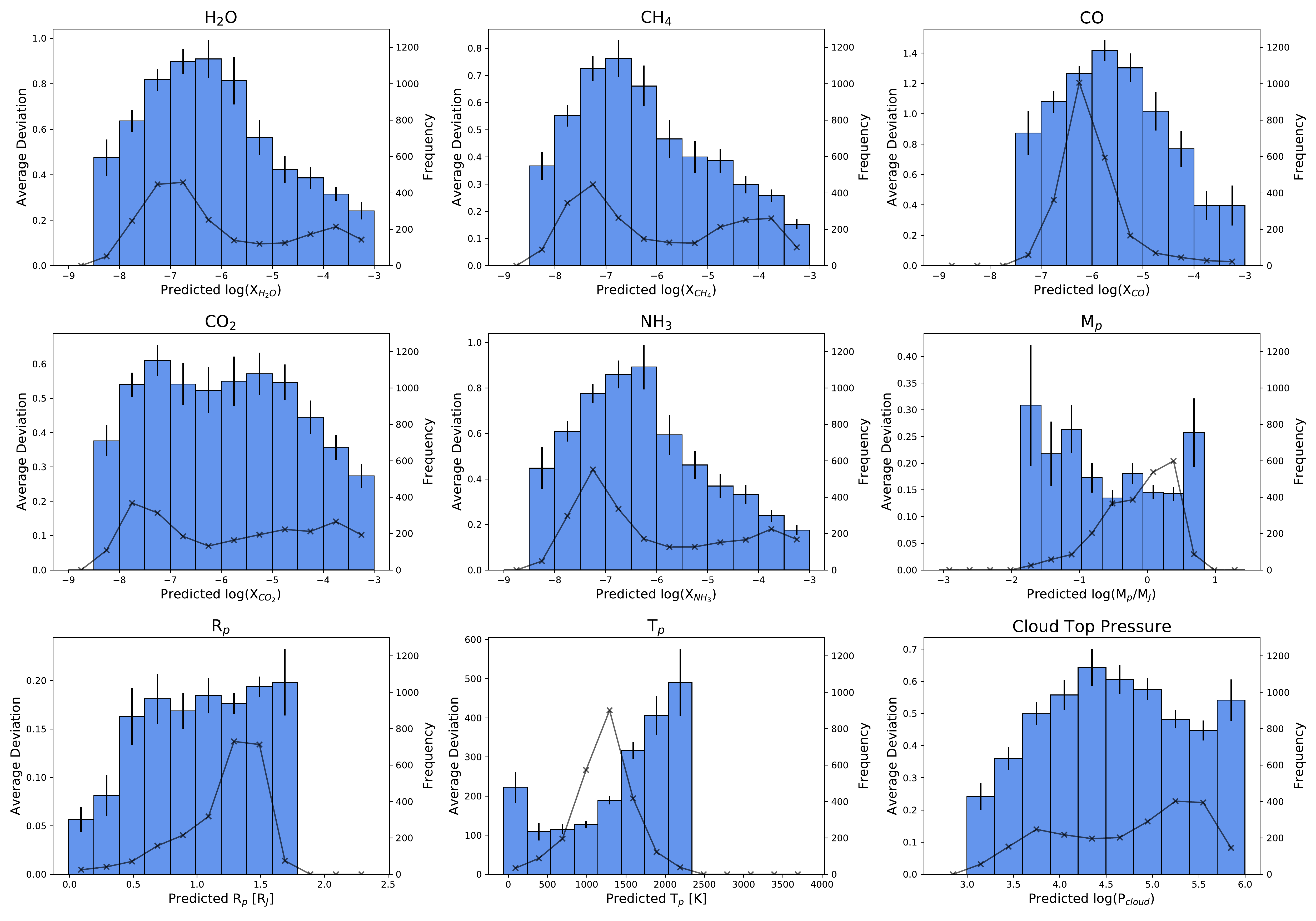}
    \caption{Distribution of average deviation from the ground truth at each prediction level for different AMPs. The deviation for each AMP is presented in actual unit. The black curve represents the frequency of model predictions for that bin. Bins are sampled at equal width with sample size less than 20 ignored. Note that M$_p$ is defined as log(M$_p$/M$_J$) for better visualisation. The model generally performs better when R$_p$ and T$_p$ is small. Whilst for M$_p$, the model struggles with extreme cases, i.e. very light or heavy planets, this could be due to lack of examples in those regions.} 
    \label{fig:all_deviation}
\end{figure}


\begin{figure}
    \centering
    \includegraphics[width=\columnwidth]{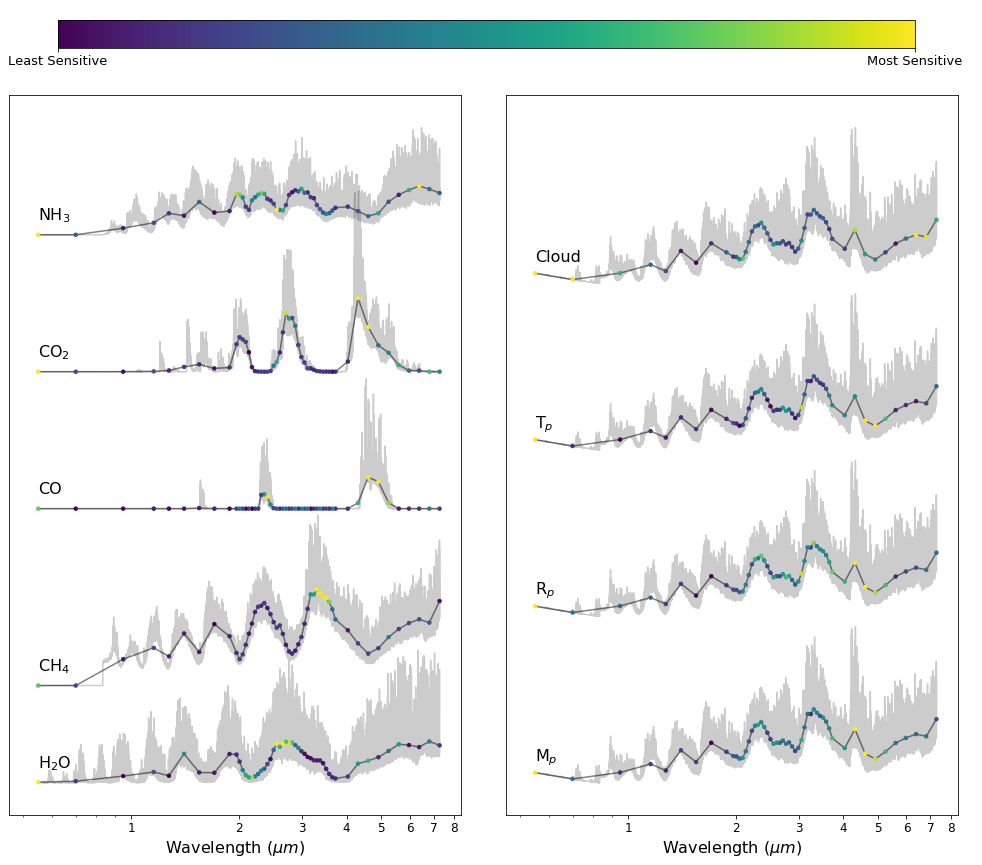}
    \caption{Sensitivity map for the MLP model. For each AMP, we visualize the sensitivity of the model to each wavelength w.r.t. predicting it. Yellow denotes the wavelength to which the model is most sensitive and black the one to which it is the least. The MLP model is more sensitive to the photometric points in most cases, which could be linked to obtaining a baseline for the model. For non-gaseous AMPs, in contrast to the CNN and LSTM (see below) models, the MLP focuses more on other parts of the spectrum than the region 2-3 $\mu m$. This hints that the MLP captures different aspects of the underlying physics than the  other 2 models. This is not surprising, as both CNN and LSTM architectures encode more local structure constraints than MLPs.} 
    \label{fig:mlp_sensi}
\end{figure}
\begin{figure}
    \centering
    \includegraphics[width=\columnwidth]{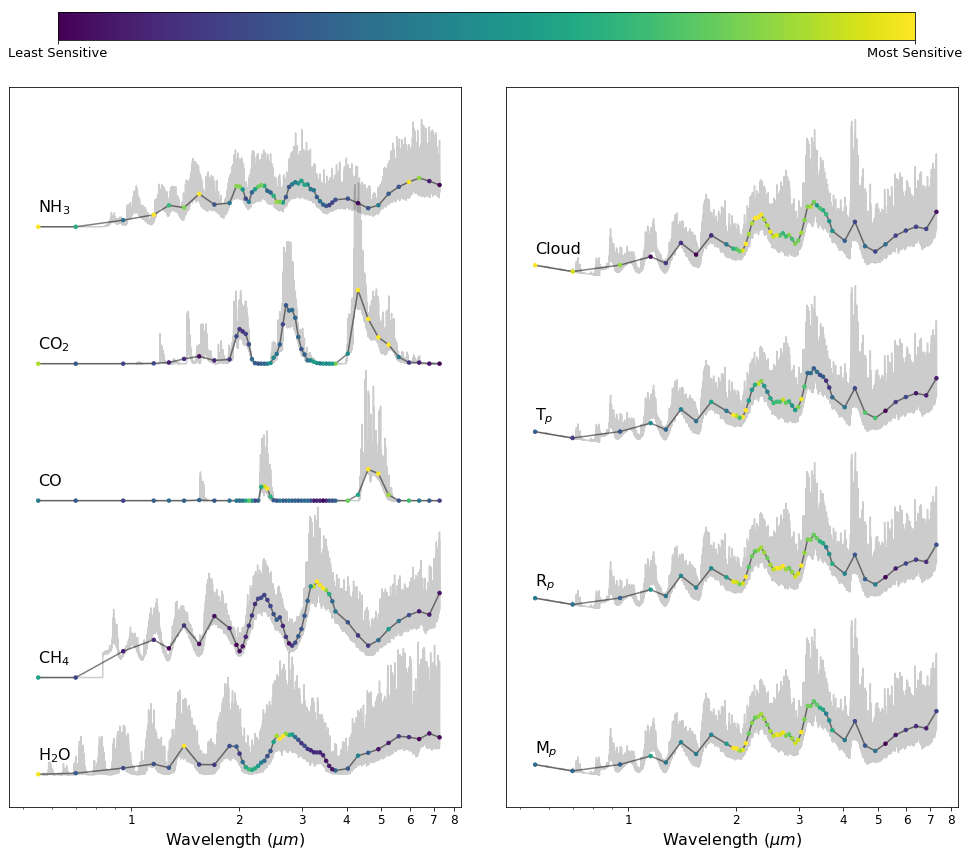}
    \caption{Sensitivity map for the LSTM model. For each AMP, we visualize the sensitivity of the model to each wavelength w.r.t. predicting it. Yellow denotes the wavelength to which the model is most sensitive and black the one to which it is the least. Similar to the CNN model, the LSTM model focuses in the region 2-3 $\mu m$ for non-gaseous AMPs.} 
    \label{fig:lstm_sensi}
\end{figure}
{\small
		\bibliographystyle{aasjournal}
		\bibliography{main} 
	}

\end{document}